\documentclass[aps,preprint,prd,showpacs,nofootinbib]{revtex4}

\usepackage{latexsym}
\usepackage{amsmath}
\usepackage{graphicx}
\usepackage{subfigure}
\usepackage{dcolumn}
\usepackage{bm}
\usepackage{amssymb}
\usepackage{latexsym}
\usepackage{ulem}
\usepackage{color}
\usepackage[colorlinks,linkcolor=magenta,anchorcolor=cyan,citecolor=blue]{hyperref}

\def\be{\begin{equation}}
\def\ee{\end{equation}}
\def\ba{\begin{eqnarray}}
\def\ea{\end{eqnarray}}

\def\nn{\nonumber}
\def\lf{\left}
\def\rt{\right}

\begin{document}

\title{Primordial perturbations with pre-inflationary bounce }

\author{Yong Cai$^{1}$\footnote{caiyong13@mails.ucas.ac.cn}}
\author{Yu-Tong Wang$^{1}$\footnote{wangyutong12@mails.ucas.ac.cn}}
\author{Jin-Yun Zhao$^{1}$\footnote{zhaojinyun15@mails.ucas.ac.cn}}
\author{Yun-Song Piao$^{1,2}$\footnote{yspiao@ucas.ac.cn}}

\affiliation{$^1$ School of Physics, University of Chinese Academy of
Sciences, Beijing 100049, China}


\affiliation{$^2$ Institute of Theoretical Physics, Chinese
Academy of Sciences, P.O. Box 2735, Beijing 100190, China}

\begin{abstract}

Based on the effective field theory (EFT) of nonsingular
cosmologies, we build a stable model, without the ghost and
gradient instabilities, of bounce inflation (inflation is preceded
by a cosmological bounce).  We perform a full simulation
for the evolution of scalar perturbation, and find that the
perturbation spectrum has a large-scale suppression (as expected),
which is consistent with the power deficit of the cosmic
microwave background (CMB) TT-spectrum at low
multipoles, but unexpectedly, it also shows itself one marked
lower valley, which actually provides a better fit to the dip at
multipole $l\sim 20$. The depth of valley is relevant with the
physics around the bounce scale, which is model-dependent.

\end{abstract}

\maketitle

\section{Introduction}

Inflation
\cite{Guth:1980zm}\cite{Linde:1981mu}\cite{Albrecht:1982wi}\cite{Starobinsky:1980te}
is the current paradigm of early universe. It predicts nearly
scale-invariant scalar perturbation, which is consistent with the
cosmic microwave background (CMB) observations \cite{Ade:2015xua}\cite{Ade:2015lrj}, as well as
the gravitational waves (GWs). However, it is not the final story
of the early universe. As pointed out by Borde, Vilenkin and Guth
\cite{Borde:1993xh}\cite{Borde:2001nh}, inflation is
past-incomplete, and ``\textit{inflationary models require physics
other than inflation to describe the past boundary of the
inflating region of spacetime.}" \cite{Borde:2001nh}.

This past-incompletion (singularity) of inflation has inspired
radical alternatives to inflation, e.g.,
\cite{Khoury:2001wf}\cite{Cai:2007qw}\cite{Piao:2003ty}\cite{Creminelli:2010ba}.
However, how to make the inflation happen in a past-complete
scenario is also a noteworthy issue. In certain sense, this
actually requires that the pre-inflationary phase should be
past-complete. One possibility is that it is slow contracting, so
that the infinite past is complete Minkowski spacetime. In such a
scenario, a nonsingular bounce preceding inflation must occur
(so-called the bounce inflation scenario) \cite{Piao:2003zm}.

Recently, the Planck collaboration
\cite{Ade:2013sjv}\cite{Ade:2015hxq} have observed the power
deficit of CMB TT-spectrum at large scale. This might be a hint of
the pre-inflationary physics, which happens around $\sim 60$
efolds, e.g., \cite{Cai:2015nya}. The idea of bounce inflation
accounted for not only the power deficit on large angular scales
\cite{Piao:2003zm}\cite{Liu:2013kea}\cite{Biswas:2013dry}, but
also a large dipole power asymmetry
\cite{Liu:2013kea}\cite{Liu:2013iha} in the CMB fluctuation. Thus
we conjectured that the physics hinted by the CMB anomalies might
be relevant with the pre-inflationary bounce, see also
\cite{Falciano:2008gt}\cite{Mielczarek:2008pf}\cite{Xia:2014tda}\cite{Liu:2014tda}\cite{Qiu:2015nha}\cite{Wan:2015hya}\cite{Li:2016awk}\cite{Ni:2017jxw}.

In physical time, the equation of motion  of scalar perturbation
$\zeta$ is \be \ddot{\zeta}_k+\lf(3H+{\dot{Q}_s\over
Q_s}\rt)\dot{\zeta}_k+c_s^2{k^2\over
a^2}\zeta_k=0\,.\label{eom-zeta} \ee Generally, $Q_s\sim
\epsilon_{cont}=const.\gg 1$ for the contraction, while $Q_s\sim
\epsilon_{inf}<1$ for the inflation, where
$\epsilon=-{\dot{H}/H^2}$. Thus $Q_s$ inevitably shows itself a
jumping around the nonsingular bounce, even if this phase lasts
shortly enough. Previous studies neglected the effect of $Q_s$ on
the perturbation spectrum, since this effect is ambiguous without
a fully stable (without the ghost and gradient instabilities)
nonsingular bounce. Recently, with the effective field theory
(EFT) of nonsingular cosmologies
\cite{Cai:2016thi}\cite{Creminelli:2016zwa}\cite{Cai:2017tku}, we
have been able to stably manipulate the bounce
\cite{Cai:2017dyi}\cite{Kolevatov:2017voe}, see also
\cite{Cai:2017dxl}\cite{Ijjas:2016vtq}.  This impels us to
reconsider the relevant issue.


In this paper, inspired by
\cite{Cai:2016thi}\cite{Creminelli:2016zwa}\cite{Cai:2017dyi}\cite{Kolevatov:2017voe},
we build a fully stable model of bounce inflation, in which
initially the universe is in the ekpyrotic contraction.  By
numerically solving Eq. (\ref{eom-zeta}), we find that the
pre-inflationary bounce not only brings the power deficit of the
CMB TT-spectrum at low multipoles (as expected in
\cite{Piao:2003zm}\cite{Liu:2013kea}), but unexpectedly, also
provides a better explanation to the dip at multipole $l\sim
20$ hinted by Planck \cite{Ade:2015lrj}.

\section{The Lagrangian }\label{pbounceinf}

Recently, it has been found that the nonsingular cosmological
models usually suffer from the ghost or gradient instabilities
($c_s^2<0$) \cite{Libanov:2016kfc}\cite{Kobayashi:2016xpl}, 
see also \cite{Kolevatov:2016ppi}\cite{Akama:2017jsa}. Based on
the EFT of nonsingular cosmologies
\cite{Cai:2016thi}\cite{Creminelli:2016zwa}\cite{Cai:2017tku},
this No-go result has been clearly illustrated. The cubic Galileon
interaction $\sim\Box\phi$ in Horndeski theory
\cite{Horndeski:1974wa}\cite{Deffayet:2011gz}\cite{Kobayashi:2011nu}
only moves the period of $c_s^2<0$ to the outside of bounce phase,
but cannot dispels it completely
\cite{Ijjas:2016tpn}\cite{Easson:2011zy}.  It has been
found first in \cite{Cai:2016thi}\cite{Creminelli:2016zwa} that
the operator $R^{(3)}\delta g^{00}$ in EFT could play significant
role in curing the gradient instability of scalar perturbation. 
Recently, we have built fully stable cosmological bounce models in
Ref. \cite{Cai:2017dyi} by applying the covariant $L_{R^{(3)}\delta
g^{00}}$.

We follow Ref. \cite{Cai:2017dyi}, and after defining
$\phi_\mu=\nabla_\mu\phi$, $\phi^\mu=\nabla^\mu\phi$,
$\phi_{\mu\nu}=\nabla_\nu\nabla_\mu\phi$,
$X=\phi_{\mu}\phi^\mu$ and $\Box\phi=\phi^\mu_{~\mu}$, write
the effective Lagrangian of nonsingular bounce inflation
as ($\phi$ is set dimensionless) \ba\label{action01}  L\sim & &
\underbrace{{M_p^2\over
2}R-{M_p^2\over 2}X-V(\phi)} \nn\\ & & \text{Contraction}+\text{Inflation}\nn\\
&+& \quad\quad\underbrace{{\tilde
P}(\phi,X)}\quad\quad\quad\quad\quad\quad +\underbrace{L_{\delta
g^{00} R^{(3)}}} \quad\quad+L_{\delta K \delta g^{00} }\,, \\
& & \text{(Ghost free) Bounce} \quad\quad\,\, \text{Removing} ~
c_s^2<0 \nn\ea where
\ba \label{covaaction1} L_{\delta g^{00} R^{(3)}}& =& {f_1(\phi)\over 2}\delta g^{00} R^{(3)}\nn\\
&=& {f\over 2}R -{X\over 2} \int f_{\phi\phi}d \ln X
-\lf(f_\phi+\int {f_\phi \over 2}d\ln X\rt)\Box\phi \nn\\
& \, & + {f\over 2X}\lf[\phi_{\mu\nu}\phi^{\mu\nu}-(\Box
\phi)^2\rt]-{f-2Xf_X\over
    X^2}\lf[\phi^\mu\phi_{\mu\rho}\phi^{\rho\nu}\phi_\nu-(\Box
\phi)\phi^\mu\phi_{\mu\nu}\phi^\nu\rt] \,, \\ L_{\delta K \delta
g^{00} }&=& {g_1(\phi)\over2}\delta K\delta g^{00}
\nn\\
&=&{g\over
2}{1\over\sqrt{-X}}\lf({\phi^\mu\phi_{\mu\nu}\phi^\nu\over X}-\Box
\phi \rt) -{3\over2}g H\,,\\ f &=& f_1(\phi)\lf[ 1+{X\over
f_2(\phi)} \rt],\quad\quad g=g_1(\phi)\lf[1+{X\over
f_2(\phi)}\rt]\,,\ea with $f_2= {X\over \delta
g^{00}-1}={\dot \phi}^2(t)$,  $R^{(3)}\delta g^{00}$ and
$\delta K\delta g^{00}$ being the EFT operators ($R^{(3)}$ is the
3-dimensional Ricci scalars on the spacelike hypersurface). We
briefly review the EFT of nonsingular cosmologies in Appendix \ref{appA},
see (\ref{eft_action}) for the definition of $\delta g^{00}$ and
$\delta K$. Though $L_{\delta g^{00} R^{(3)}}$ has the higher
order of the second order derivative of $\phi$, it is Ostrogradski
ghost-free \cite{Langlois:2015cwa}\cite{Langlois:2015skt}. 
Additionally, ${L_{\delta g^{00} R^{(3)}}}$ and $L_{\delta K \delta
g^{00}}$ do not affect the cosmological background.

\section{A stable model of bounce inflation} \label{sec-model}

\subsection{Background}

\begin{figure}[htbp]
    \includegraphics[scale=2,width=0.6\textwidth]{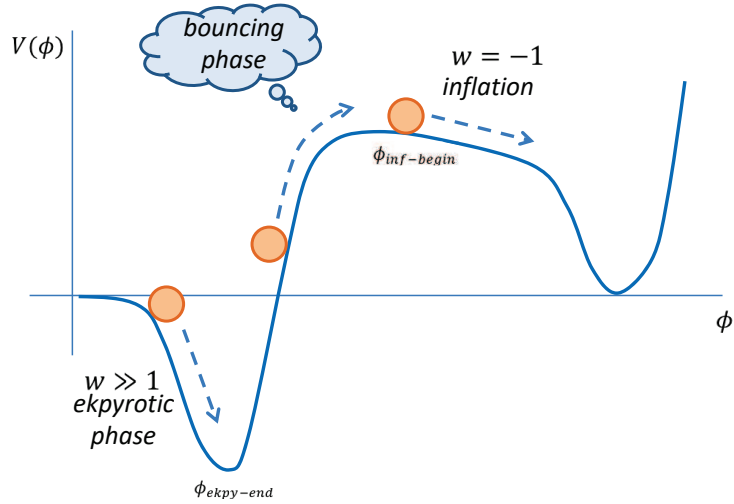}
    \caption{~~A sketch of the bounce inflation scenario. }
    \label{sketch}
\end{figure}

A sketch of the bounce inflation scenario is plotted in Fig.
\ref{sketch}.  We will show how to build its stable model with
the Lagrangian (\ref{action01}).

As a specific model, we set \be \label{tildeP} {\tilde
P}(\phi,X)={\alpha_0\over \lf(1+(\phi/\lambda_1)^2
\rt)^2}M_p^2\,X/2+ {\beta_0\over\lf(1+(\phi/\lambda_1)^2
\rt)^2}X^2/4, \ee \be \label{V} V(\phi)=-{V_0\over 2}
e^{\sqrt{2\over q}\phi}\lf[1-\tanh\lf( {\phi\over \lambda_2}\rt)
\rt]+{\Lambda\over 2} \lf(1-\lf({\phi\over \lambda_3}\rt)^2 \rt)^2
\lf[1+\tanh\lf( {\phi\over \lambda_2}\rt) \rt]\,, \ee with the
positive constants $\lambda_{1,2,3}$ and
$q,\alpha_0,\beta_0$ being  dimensionless.  We have
${\tilde P}(\phi,X)\neq 0$ only around $\phi\simeq 0$
\cite{Buchbinder:2007ad}\cite{Battarra:2014tga}\cite{Koehn:2015vvy},
while ${\tilde P}(\phi,X)= 0$ for $|\phi|\gg \lambda_1$.


Thus we have \ba \label{eqH} 3 H^2 M_p^2 &=& \lf[1-{\alpha_0\over
\lf(1+(\phi/\lambda_1)^2 \rt)^2}\rt]M_p^2\,\dot{\phi }^2/2{+{3\beta_0\over \lf(1+(\phi/\lambda_1)^2 \rt)^2} }\dot{\phi
}^4/4+V(\phi) \,,
\\
\dot{H} M_p^2 &=& -\lf[1-{\alpha_0\over \lf(1+(\phi/\lambda_1)^2
\rt)^2}\rt]M_p^2\,\dot{\phi }^2/2{-{\beta_0\over
\lf(1+(\phi/\lambda_1)^2 \rt)^2}\dot{\phi }^4/2}\,.\label{dotH} \ea
In infinite past, the universe is almost Minkowski, which will
experiences the ekpyrotic contraction. In the ekpyrotic phase
($\phi\ll -\lambda_1$  and $-\lambda_2$), we have ${\tilde P}=0$ and $V_{ekpy}= -V_0
e^{\sqrt{2\over q}\phi}$ ($q\ll 1$). Thus we could write Eqs. (\ref{eqH}) and (\ref{dotH})
as \be  3H^2= \dot{\phi }^2/2 -{V_0\over M_p^2}e^{\sqrt{2\over q}\phi},\quad\quad
\dot{H}=-\dot{\phi }^2/2. \label{ekpy}\ee
By solving (\ref{ekpy}), we have \be \label{dotphi}
a\sim (-t)^{1/\epsilon}\,,\quad\quad \dot{\phi}=\sqrt{2\over \epsilon}(-t)^{-1}\,, \ee and \be
\label{phit} \phi(t)=\sqrt{2\over \epsilon }\ln\lf[ {\sqrt{\epsilon-3}
\over \epsilon {\sqrt{V_0}/M_p}}(-t)^{-1}\rt]\,, \ee where
$\epsilon=-\dot{H}/H^2=1/q\gg 1$, which suggests $H=-{
\epsilon}^{-1}(-t)^{-1}$.

When $\phi\simeq \lambda_1$, we could have \be \dot{H} \simeq
\lf({\alpha_0\over 4}-{\beta_0\dot{\phi }^2\over 
4M_p^2}-1\rt)\,\dot{\phi }^2/2>0\,,\ee the nonsingular bounce
will occur. While after $\phi\gg \lambda_1, \lambda_2$, the
field $\phi$ will be canonical (${\tilde P}=0$) again. We have \be
3H^2= \dot{\phi }^2/2+{\Lambda\over
M_p^2}\lf(1-\lf({\phi\over \lambda_3}\rt)^2 \rt)^2,\quad\quad
\dot{H}=-\dot{\phi }^2/2. \label{inf}\ee Thus the slow-roll
inflation will occur. Actually, after the nonsingular
bounce, the Lagrangian (\ref{action01}) will reduce to $L\sim
{M_p^2}R/2-{M_p^2}X/2-V_{inf}$ with $V_{inf}$ being the potential
of slow-roll inflation.

\begin{figure}[htbp]
\subfigure[~~$\phi$]{\includegraphics[width=.48\textwidth]{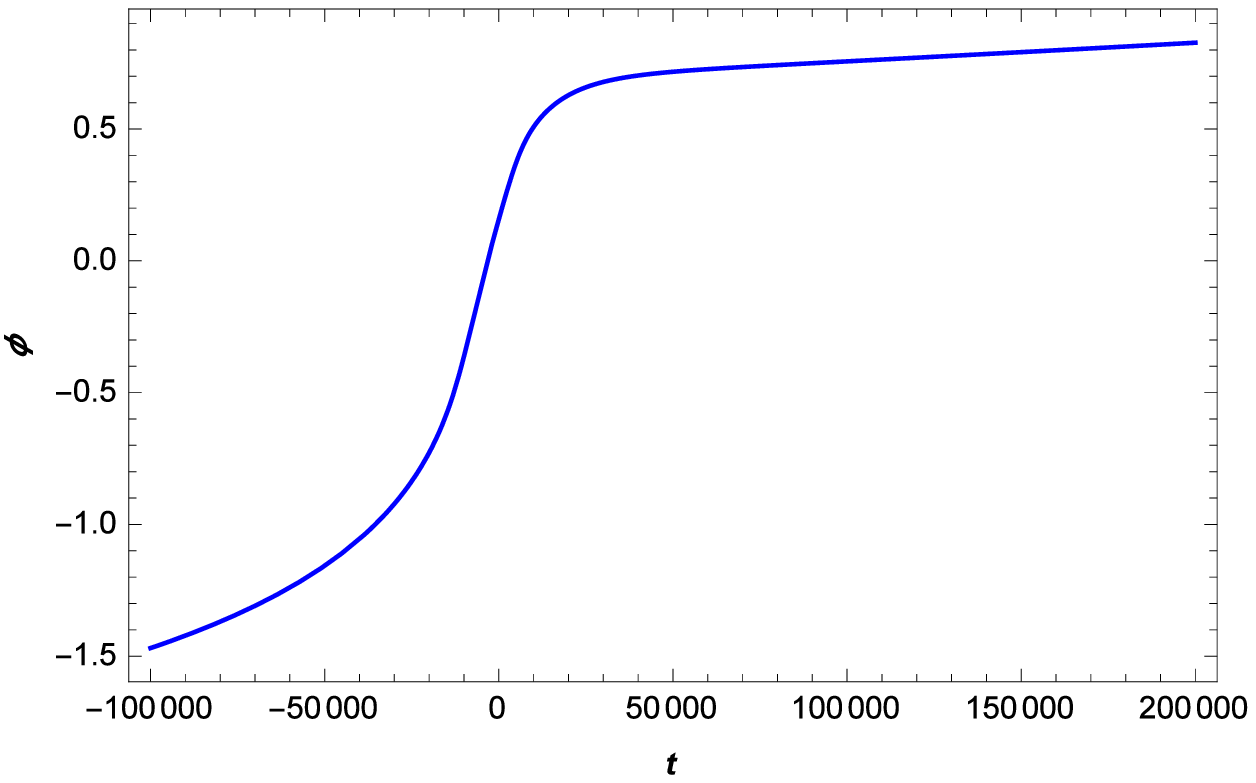}
}
\subfigure[~~$a$]{\includegraphics[width=.47\textwidth]{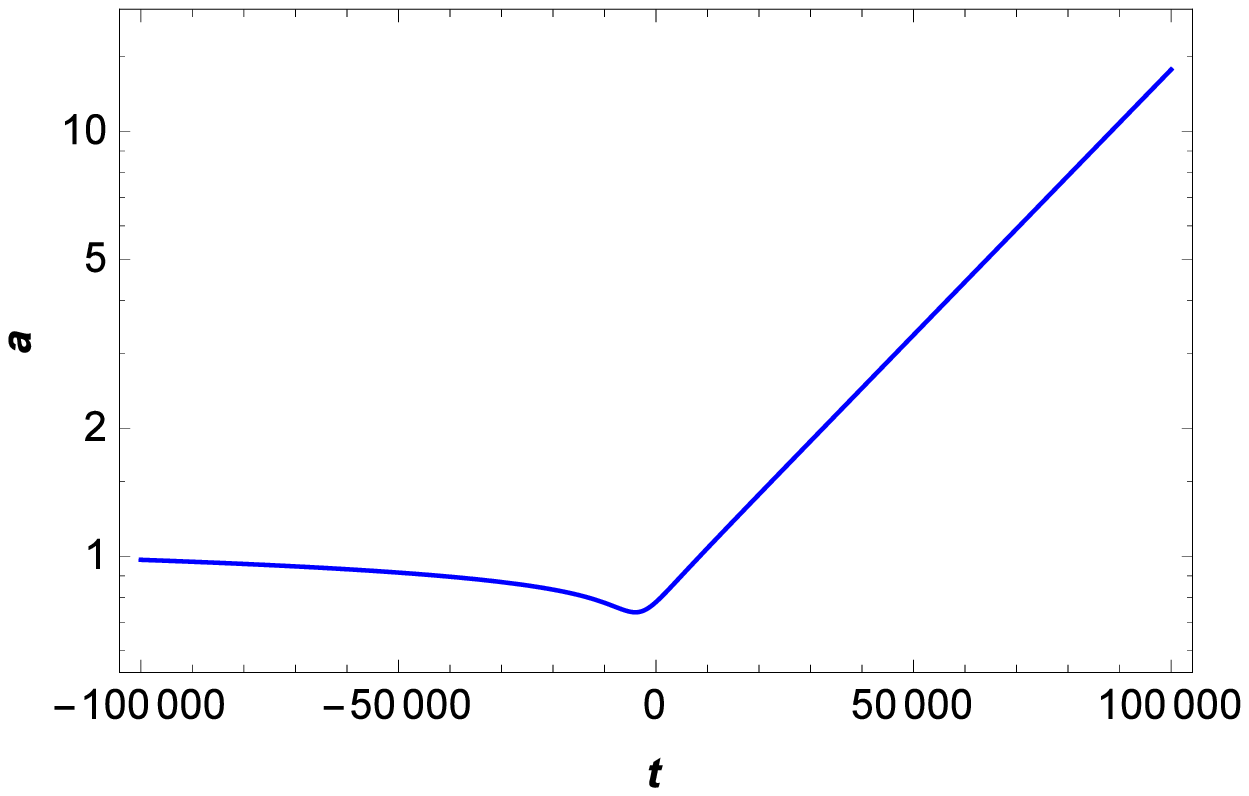}
} \subfigure[~~$10^5\cdot
H/M_p$]{\includegraphics[width=.48\textwidth]{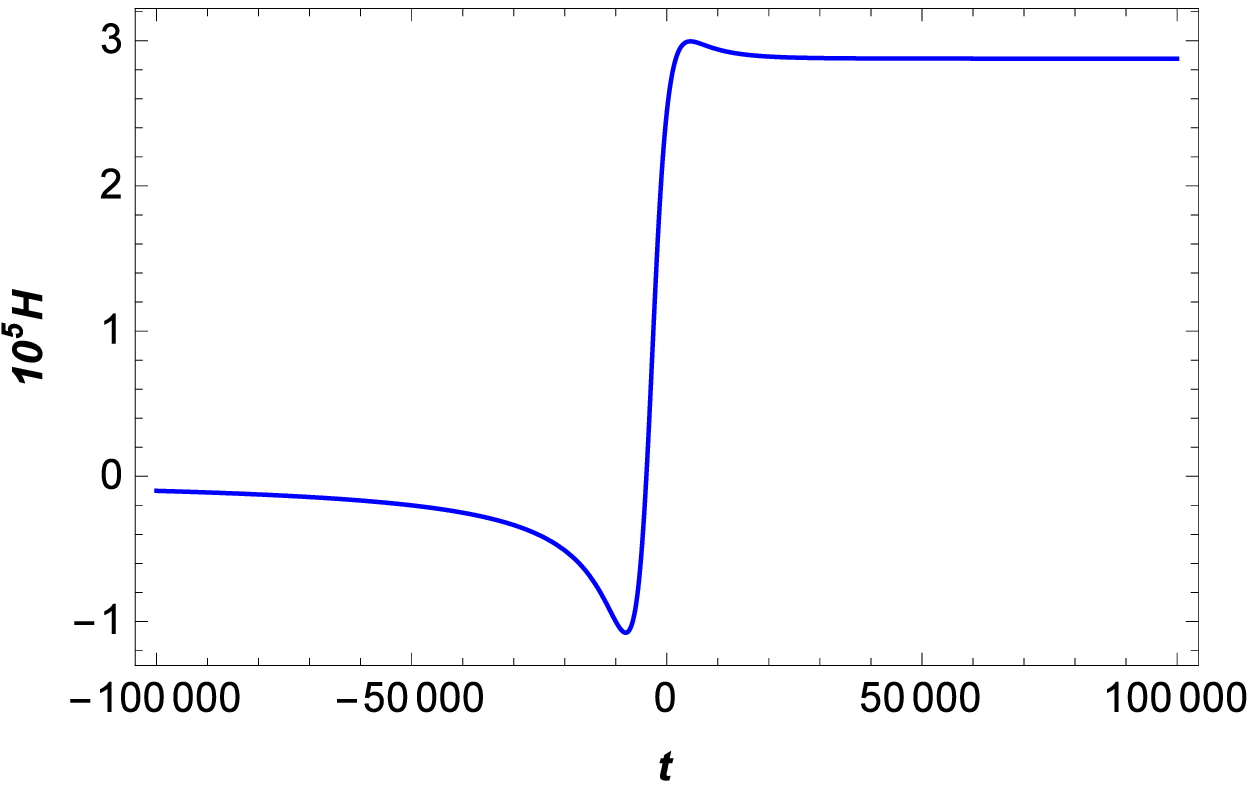} }
\subfigure[~~$\epsilon=-\dot{H}/H^2$]{\includegraphics[width=.48\textwidth]{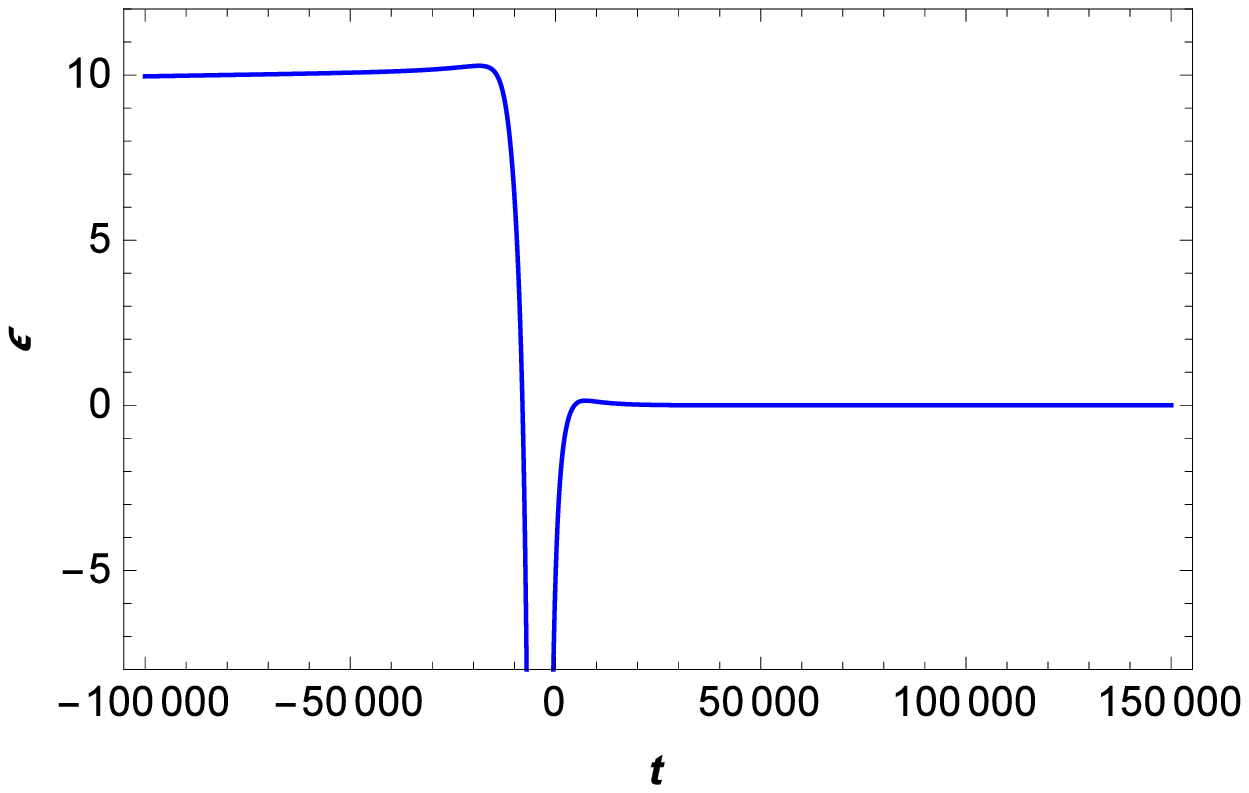}
} \caption{The background evolution of our model
with $\alpha_0=20$, $\beta_0=5\times10^9$, $\lambda_1=0.224$,
$\lambda_2=0.0667$, $\lambda_3=12$, $V_0=5\times10^{-9}M_p^4$,
$q=0.1$, $ \Lambda=2.5\times10^{-9}M_p^4$.} \label{fig01}
\end{figure}

We plot the background evolution in Fig. \ref{fig01} with
$\alpha_0=20$, $\beta_0=5\times10^9$, $\lambda_1=0.224$,
$\lambda_2=0.0667$, $\lambda_3=12$, $
V_0=5\times10^{-9}M_p^4$, $q=0.1$, $
\Lambda=2.5\times10^{-9}M_p^4$. The initial values are set by
(\ref{dotphi}) and (\ref{phit}).

\subsection{Simulation for the  scalar perturbation spectrum}

In unitary gauge $\delta\phi=0$, the quadratic action of scalar
perturbation $\zeta$ for (\ref{action01}) is (see Appendix \ref{appA} and
also our \cite{Cai:2016thi}) \be S_{\zeta}^{(2)}=\int a^3
Q_s\lf(\dot{\zeta}^2-c_s^2 {(\partial\zeta)^2\over a^2} \rt)d^4x
\,,\label{scalar-action} \ee in which \be Q_s={2{\dot
\phi}^4{\tilde P}_{XX}-M_p^2{\dot H}\over \gamma^2}+{3}\lf({
g_1\over 2\gamma M_p}\rt)^2,\ee \be c_s^2Q_s= {{\dot
c}_3\over a}-M_p^2,\quad\quad c_3={a{ M_p^2}\over
\gamma}\lf(1+{2f_1\over M_p^2}\rt), \label{cs2} \ee  with
$\gamma=H+{g_1\over 2M_p^2}$.


The stabilities require $Q_s>0$ and
$c_s^2>0$. Generally, $Q_s>0$ can be obtained by applying ${\tilde P}(\phi, X)$.
While around
the bounce point $H\simeq 0$, \be c_s^2\sim -{\dot
\gamma}\lf(1+{2f_1\over M_p^2}\rt){+{2{\dot f}_1 \gamma\over
M_P^2}-\gamma^2}. \ee We will have $c_s^2>0$ only for $2f_1<-{M_p^2}$, as has
been clarified in Refs. \cite{Cai:2016thi}\cite{Cai:2017tku}. Thus
the gradient instability ($c_s^2<0$) is cured by $L_{\delta g^{00}
R^{(3)}}$, since if $f_1\equiv0$, we have $c_s^2\sim -{\dot \gamma} -\gamma^2<0$
around the bounce point. Here, we always could set $c_s^2\sim
{\cal O}(1)$ with a suitable $f_1(\phi)$ (see also
\cite{Cai:2017tku}) satisfying \be {2f_1(\phi)}={
\gamma\over a}\int a\lf(Q_s c_s^2 +M_p^2\rt)dt-M_p^2.
\label{m4tilde}\ee


In conformal time $\eta=\int dt/a$, the motion equation of $\zeta$ is \ba u''+\lf({c}_{s}^2
k^2-{z_s''\over z_s} \rt)u=0\,, \label{eom-us} \ea where
$u=z_s\zeta$ and $z_s=\sqrt{2a^2 Q_s}$. In infinite past, the universe is almost
Minkowski, and will come through the ekpyrotic phase. The
perturbation modes have the wavelength $\lambda\simeq 1/k\ll
\sqrt{z_s/z_s''}$ and $c_s^2=1$. Thus the initial state of the
perturbation is \be \label{initial-state} u\simeq {1\over
\sqrt{2k}}e^{-ik\eta}\,. \ee The perturbation modes will pass
through the ekpyrotic phase, the bounce phase and the inflation
phase, sequentially. The resulting spectrum $P_\zeta$ of $\zeta$ (at $-k\eta\ll 1$) is
\be P_{\zeta}={k^3\over 2\pi^2}|\zeta|^2\,. \ee


In physical time, the motion equation of $\zeta$ is
(\ref{eom-zeta}). In the ekpyrotic phase, $z_s\sim a\sim
(-\eta)^{1\over \epsilon_{ekpy}-1}$, since $Q_s\sim
\epsilon_{ekpy}=const.\gg 1$. While in the inflationary phase,
$\epsilon_{inf}<1$. This suggests that $Q_s$ (or $z_s\sim
a\sqrt{Q_s}$) will show itself a jumping around the nonsingular
bounce, which will inevitably affect $P_\zeta$. Whether the
jumping of $Q_s$ is gentle or not is model-dependent. We will
simulate its effect on $P_\zeta$ by numerically solving Eq.
(\ref{eom-zeta}), with $c_s^2=1$ set by
Eq.(\ref{m4tilde}).

It should be mentioned that if $g_1=0$ ($L_{\delta K \delta g^{00}
}$ is absent), we will have $\gamma=H=0$ at the bounce point and
$Q_s\sim 1/\gamma^2$ is divergent, see (\ref{cs2}), so that Eq.
(\ref{eom-zeta}) is singular. Here, in order to avoid it, we apply
$g_1(\phi)$, see also \cite{Cai:2017tku}.

\begin{figure}[htbp]
    \includegraphics[scale=2,width=0.6\textwidth]{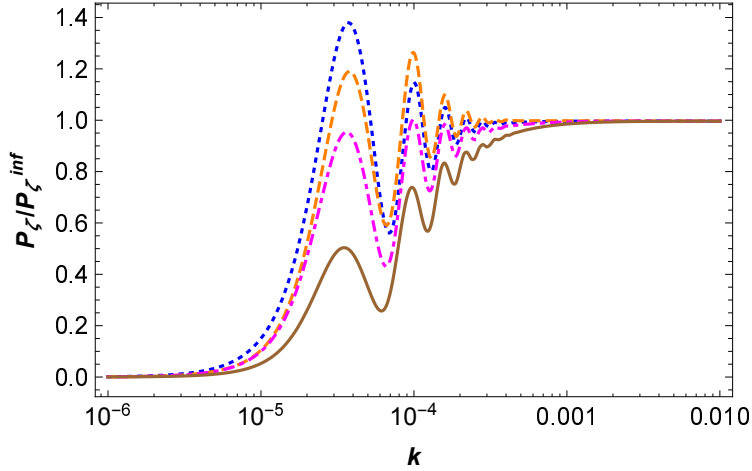}
    \caption{~~$P_{\zeta}/P_{\zeta}^{inf}$ with background set by
        Fig. \ref{fig01}, where the \{solid brown, dotdashed magenta,
        dashed orange, blue dotted\} curves correspond to ${\cal
            A}_Q=\{3,\,3,\,3,\,3\}$, ${\cal B}=\{1.3,\,1.8,\,2,\,3\}$,
        $t_*=\{4,\,4,\,2.5,\,4\}\times 10^4$, respectively. }
    \label{multi-power}
\end{figure}

Without loss of generality, we set \be Q_s={\cal
A}_Q\lf[{\cal B}-\tanh\lf( {t\over t_*}\rt) \rt]\,,\label{Qs} \ee
which requires \be g_1(\phi(t))=-\frac{2 H M_p^2 Q_s-2\sqrt{3 H^2
M_p^6 Q_s+M_p^4 \left(3 M_p^2-Q_s\right)
   \left(\dot{H} M_p^2-2  \dot{\phi }^4{\tilde P}_{XX}\right)}}{Q_s-3
   M_p^2}\,
\ee in Lagrangian (\ref{action01}), see (\ref{cs2}). We plot the
spectrum $P_\zeta$ of scalar perturbation in Fig.
\ref{multi-power} for the background in Fig. \ref{fig01} and the
different values of ${\cal B}$ and $t_*$, where
$P_{\zeta}^{inf}={H_{inf}^2\over 8Q_s^{inf}\pi^2 M_p^2}\lf({k\over
{\cal H}_{inf}}\rt)^{n_s-1}$ is that of the inflation,
with $Q_s^{inf}$ being the value of $Q_s$ during inflation,
$n_s-1\simeq 0$ (but is slightly red). The evolutions of $Q_s$,
$g_1$ and $|\zeta|$ with respect to $t$, respectively,
are plotted in Figs. \ref{fig02} and \ref{fig03} of Appendix \ref{appB}.

As expected in \cite{Piao:2003zm}, $P_\zeta$ shows itself a
large-scale cutoff, but is flat (with a damped oscillation) at
small scale. However, due to the  step-like evolution of
$Q_s$, the peaks and valleys of the oscillations are obviously
pulled lower. Actually,  after the nonsingular bounce, with
Eq. (\ref{eom-zeta}), we shortly have the effective Hubble
parameter \be H_{inf}^{eff}=H_{inf}+{{\dot Q}_s \over
3Q_s}<H_{inf}, \label{Heff}\ee since ${\dot Q}_s< 0$, see Figs.
\ref{fig02}(b) and \ref{fig03}(b) in Appendix \ref{appB}. Thus $P_\zeta$ is
pulled lower at the corresponding scale, since $P_\zeta\sim
(H_{inf}^{eff})^2$. The change rate of $Q_s$ is relevant to the
physics of nonsingular bounce, as showed in Eq. (\ref{Qs}), so the
depth of valley pulled lower is actually model-dependent.

In Sec. \ref{sec-Template}, we will show that such a marked lower
valley at corresponding scale helps to explain the dip
around $l\simeq 20$ hinted by Planck \cite{Ade:2015lrj}.

\section{More on the spectrum}

\subsection{Analytical estimation}\label{sec-ana}

We will attempt to analytically estimate $P_\zeta$. The equation
of motion for $\zeta$ is (\ref{eom-us}). In \cite{Li:2016awk}, the
spectrum of primordial GWs has been calculated. Here, if the
effect of $Q_s$ is neglected, the calculation will be similar.

The bounce phase is the evolution with ${\dot H}>0$.  We
define that it begins and ends at $\eta_{B-}$ and $\eta_{B+}$,
respectively, at which ${\dot H}=0$. We set that $H=0$ at
$\eta_B$, which corresponds to the bounce point. Generally,
$\Delta\eta_B=\eta_{B+}-\eta_{B-}\lesssim 1/{{\cal H}}_{B+}$.


In our model (Sec. \ref{sec-model}), the contracting phase
($\eta<\eta_{B-}$) is ekpyrotic-like, $a$ is almost constant
for $\epsilon_{ekpy}\gg 1$. Considering the continuities of $a$
and $H$ at $\eta_{B-}$, we have \be a(\eta)=a_{B-}\lf[{ x\over
(\epsilon_{ekpy}-1)^{-1}{\cal H}_{B-}^{-1}}\rt]^{1\over
\epsilon_{ekpy}-1},\ee  see \cite{Li:2016awk} for the details,
where ${\cal H}_{B-}$ is the comoving Hubble parameter at
$\eta_{B-}$ and $x=\eta-\eta_{B-}+(\epsilon_{ekpy}-1)^{-1}{\cal
H}_{B-}^{-1}$. We have $z^{\prime\prime}_s/z_s=a^{\prime\prime}/a$, since $Q_s$ is constant.
Thus the solution of (\ref{eom-us}) is \be
u_k={\sqrt{\pi|x|}\over2}c_{1,1}H^{(1)}_{\nu_1}(-k x) \ee where
$\nu_1=1/2$ for $\epsilon_{ekpy}\gg 1$, and
the initial condition (\ref{initial-state}) has been used.

In the nonsingular bounce phase ($\eta_{B-}<\eta<\eta_{B+}$), $H$
should cross 0. We parameterize it as $ H={\alpha}(t-t_B)$
\cite{Cai:2007zv} with ${\alpha} M_P^2\ll 1$. We have
\be\label{bounce-a} { a}\simeq {a}_B e^{{1\over2} {
\alpha}(t-t_B)^2}\simeq { a}_B\left[1+{{\alpha}\over
2}(t-t_B)^2\right], \ee where ${a}={a}_B$ at the bouncing
point $t=t_B$. The continuities of $a$ and ${\cal H}$ at
$\eta_{B-}$ and $\eta_{B+}$ suggest ${{\cal H}}_{B+}={{\cal
H}}_{B-} + {\alpha} { a}_B^2 \lf(\eta_{B+}-\eta_{B-}\rt)$. In our
models, $|{{\cal H}}_{B-}|\lesssim {{\cal H}}_{B+}/4$,  see
Figs. \ref{fig02} and \ref{fig03} in Appendix \ref{appB}, so that we
approximately have \be {{\cal H}}_{B+}\simeq {\alpha} { a}_B^2
\Delta\eta_{B}. \label{HBalpha}\ee
Thus in this phase the equation (\ref{eom-us}) is \be
u_k^{\prime\prime}+(k^2-{\alpha} a_B^2)u_k=0. \label{Bphase}\ee
Its solution is \be 
u_k(\eta)=c_{2,1}e^{l(\eta-\eta_B)}+c_{2,2}e^{-l(\eta-\eta_B)},\ee
where $l=\sqrt{{\alpha}{ a}_B^2-k^2}$. Here, we have
neglected the effect of $Q_s$, or it is difficult to solve Eq.
(\ref{eom-us}).

In inflationary phase ($\eta\geqslant \eta_{B+}$), $Q_s^{inf}$ is
almost constant. Considering the continuities of $a$ and $\cal H$
at $\eta_{B+}$, we have \be a_{inf}(\eta)=a_{B+}\lf(-y{\cal
H}_{B+}\rt)^{1\over\epsilon_{inf}-1}, \ee where
$y=\eta-\eta_{B+}+1/{\cal H}_{B+}$, and $H_{B+}={\cal H}_{B+}/a$,
$H_{inf}\lesssim H_{B+}$. The solution of (\ref{eom-us}) is
\be
u_k={\sqrt{\pi|y|}\over2}\lf[c_{3,1}H^{(1)}_{\nu_2}(-ky)+c_{3,2}H^{(2)}_{\nu_2}(-ky)\rt]
\ee where $\nu_2={\epsilon_{inf}-3\over 2(\epsilon_{inf}-1)}$.

We have $P_\zeta$ as \be P_\zeta(k,{\cal H}_{B+},{\cal H}_{B-},\Delta\eta) \approx \frac{H_{inf}^2}{8\pi^2
Q_s^{inf}M_p^2} |c_{31}-c_{32}|^2=
P_{\zeta}^{inf}|c_{31}-c_{32}|^2~,\label{pt22} \ee where
$P_{\zeta}^{inf}=\frac{ H_{inf}^2}{8\pi^2Q_s^{inf}M_p^2}$ is that of
the slow-roll inflation. Requiring the continuities of $\zeta$ and
$\dot \zeta$, we could write the coefficients as \be \left(
  \begin{array}{ccc}
    c_{3,1} \\
    c_{3,2} \\
  \end{array}
\right)
={\cal{M}}^{(3,2)} \times  {\cal{M}}^{(2,1)}\times
\left(
  \begin{array}{ccc}
    c_{1,1} \\
    c_{1,2} \\
  \end{array}
\right), \ee see Appendix \ref{appC} for the matrices ${\cal{M}}^{(2,1)}$
and ${\cal M}^{(3,2)}$.

The effects of bounce has been encoded in ${\cal M}^{(3,2)}$ and
${\cal M}^{(2,1)}$ (or $|c_{3,1}-c_{3,2}|^2$). We approximately
have \be\label{largek} |c_{3,1}-c_{3,2}|^2\approx1-{\cal
A}\sin\lf({2k\over{\cal{H}}_{B+}}\rt)-{\cal
A}\sin\lf({2k\over{\cal{H}}_{B+}}+2k\Delta\eta_{B}\rt) \ee for
$k\gg {\cal H}_{B+}$, where \be {\cal A}={{\cal{H}}_{B+}\over
k}\lf(1-{{\alpha}{
a}_B^2\over2{\cal{H}}_{B+}}\Delta\eta_B\rt)\simeq
{{\cal{H}}_{B+}\over 2k} \ee and (\ref{HBalpha}) is used, which
suggests that on small scale $k\gg {\cal H}_{B+}$, $P_\zeta$ is
flat with a rapidly damped oscillation, its maximal oscillating
amplitude is around $k\simeq {\cal H}_{B+}$. However, if the
bounce phase lasts shortly enough, $\Delta\eta_B\ll 1/{\cal
H}_{B+}$, (\ref{largek}) will be \be
|c_{3,1}-c_{3,2}|^2\approx1-{{\cal{H}}_{B+}\over
k}\sin\lf({2k\over{\cal{H}}_{B+}}\rt).\label{largek1}\ee While on
large scale $k\ll {\cal H}_{B+}$, $P_\zeta\sim k^2$ will have a
strongly blue tilt, since \be\label{smallk}
|c_{3,1}-c_{3,2}|^2\approx
w(\Delta\eta_B)\lf({k\over{\cal{H}}_{B+}}\rt)^{2} \ee where \be
w(\Delta\eta_B)=\lf[(1-{l^2\Delta\eta_B\over2{\cal{H}}_{B+}})\cosh(l\Delta\eta_B)+{l\over2}({1\over{\cal{H}}_{B+}}-\Delta\eta_B+{l^2\over4{\cal{H}}_{B+}}\Delta\eta_B^2)\sinh(l\Delta\eta_B)\rt]^2,
\ee which is $w(\Delta\eta_B)\simeq 1$ for $\Delta\eta_B\simeq 0$.

\begin{figure}[htbp]
    \subfigure[~~$\Delta\eta=0.1/{{\cal{H}}_{B+}}$]{\includegraphics[width=.47\textwidth]{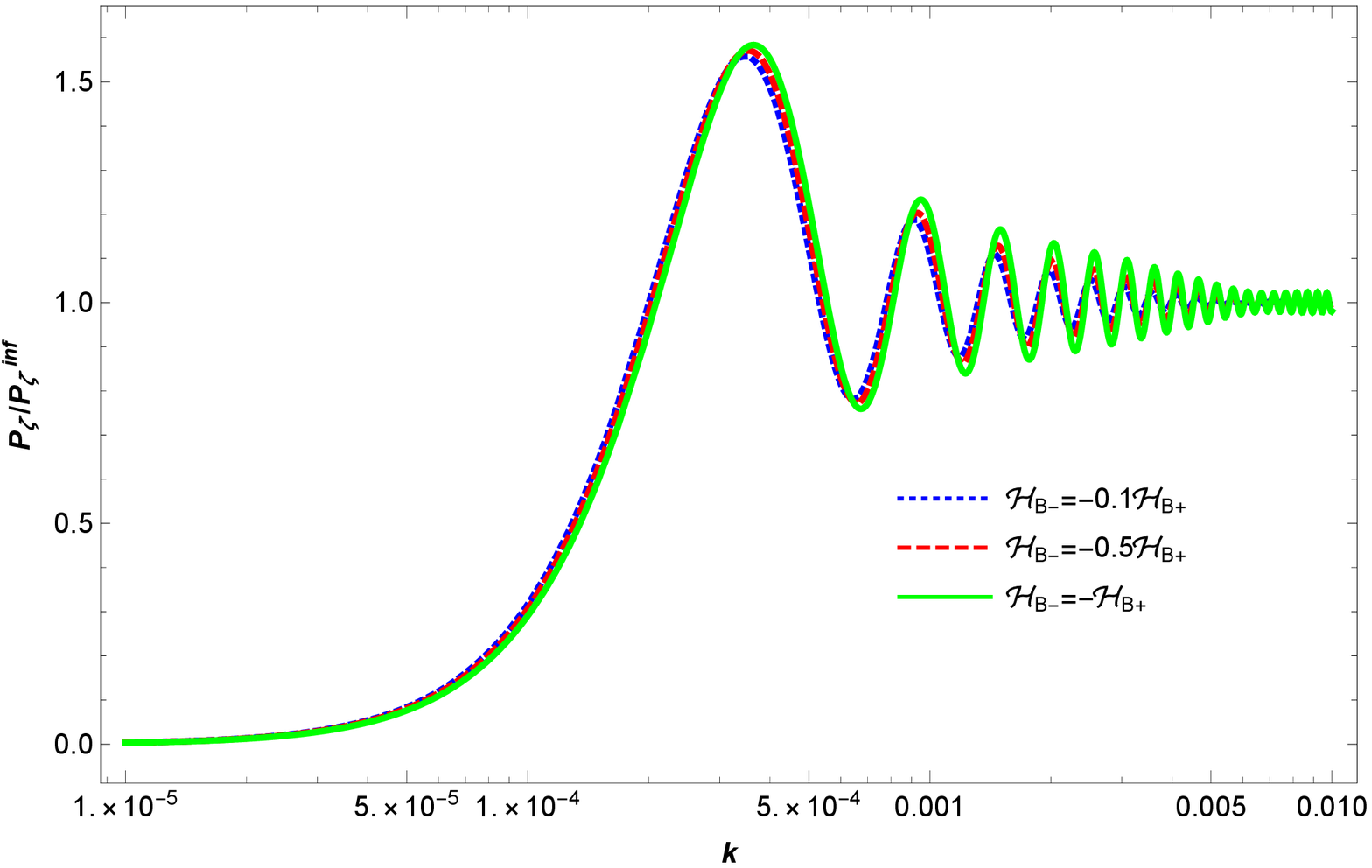} }
    \subfigure[~~$\Delta\eta=0.5/{{\cal{H}}_{B+}}$]{\includegraphics[width=.47\textwidth]{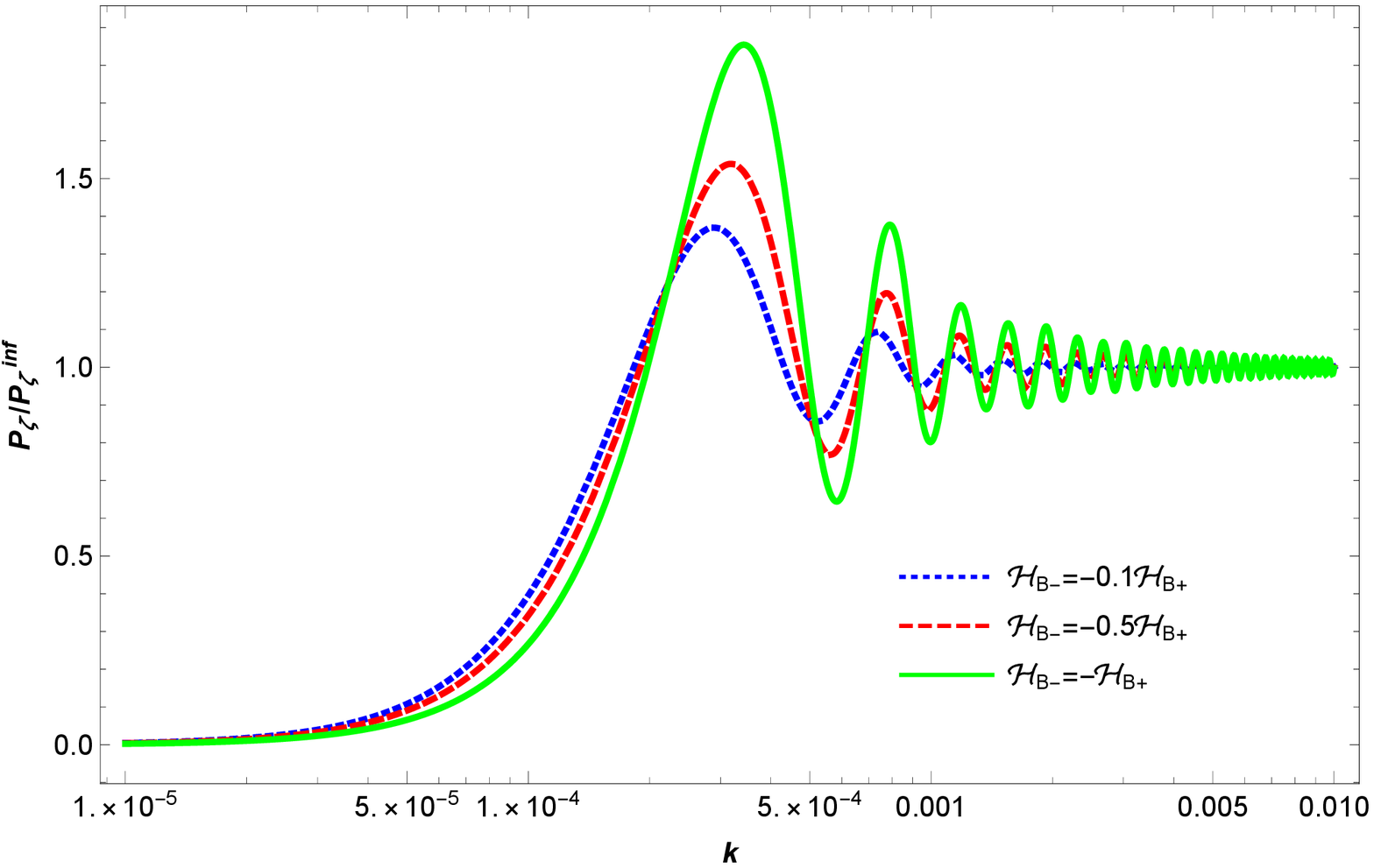} }
    \caption{ The power spectrum with different $\Delta \eta$ and different ${{\cal H}_{B-}/ {\cal H}_{B+}}$.} \label{h1h2}
\end{figure}

We plot $P_\zeta$ for (\ref{pt22}) in Figs. \ref{h1h2}
for the different values of
$\Delta\eta$ and ${\cal H}_{B-}$. We see that for $k>{\cal
H}_{B+}$, $P_\zeta\sim k^0$ but has a damped oscillation, while
for $k<{\cal H}_{B+}$, $P_\zeta\sim k^2$ shows itself a
large-scale cutoff. Thus (\ref{pt22}) is consistent with our simulation result
(see Fig. \ref{fig02} in Sec. \ref{sec-model}) well at large
and small scales, respectively.

However, since we have neglected the step-like evolution of
$Q_s$, the pull-lower around $k\simeq {\cal H}_{B+}$ in Fig.
\ref{fig02}(d) cannot be reflected in (\ref{pt22}).

%

\subsection{Template}\label{sec-Template}

To conveniently fit the observation data, a simple
\textit{``Template"} capturing the essential shape of $P_\zeta$ is
indispensable. Based on the simulation in Sec. \ref{sec-model} and
the analytical estimate in Sec. \ref{sec-ana}, we write it as \be
P_\zeta= F(k,{\cal H}_{B+},A_d,\omega_d)\cdot P^{inf}_\zeta\,, \ee
where $P_\zeta^{inf}= A_{inf}({k\over k_*})^{n_{inf}-1}$ is the
spectrum predicted by slow-roll inflation, and $A_{inf}$ is the
amplitude at the pivot scale $k_*$, $n_{inf}$ is its tilt, and \ba
\label{Template}F(k,{\cal H}_{B+}, A_d, \omega_d)&=&\Big\{1+
e^{-(k/{\cal H}_{B+})^2}\lf({k\over {\cal H}_{B+} } \rt)^{2}
\nn\\
&\,&+e^{-(k/{\cal H}_{B+})^2} -{\sin(2k/{\cal H}_{B+})\over
k/{\cal H}_{B+}} \Big\} \cdot\lf[1-A_d\cdot e^{-\omega_d({k\over
{\cal H}_{B+}}-\pi)^2} \rt]\,.\label{modfit3} \ea Here, the
parameters set $({\cal H}_{B+}, A_d, \omega_d)$ reflects the
effect of pre-inflationary bounce on the spectrum. Around
${k\gtrsim {\cal H}_{B+}}$, we have \be F(k,{\cal H}_{B+}, A_d,
\omega_d)\simeq 1-A_d\, e^{-{\cal O}(1)\omega_d}, \ee so $A_d$ and
$\omega_d$ (related with the parameter $\Delta\eta< 1/{\cal
H}_{B+}$ in Sec. \ref{sec-ana}) depict the width and depth of
valley around $k\gtrsim H_{B+}$, respectively. Here, $A_d$ is
related with the change rate of $Q_s$ (neglected in Sec.
\ref{sec-ana}). With Eq. (\ref{Heff}), we have approximately \be
A_d\simeq {2\lf|{\dot Q}_s\rt|_{max}\over 3H_{inf}Q_s} \ee noting
${\dot Q}_s<0$. In (\ref{Template}), we have \be F(k,{\cal
H}_{B+}, A_d, \omega_d)\sim 1- {\sin(2k/{\cal H}_{B+})\over
k/{\cal H}_{B+}} \ee for $k\gg {\cal H}_{B+}$, which equals to
(\ref{largek1}), while for $k\ll {\cal H}_{B+}$, we approximately
have $F(k,{\cal H}_{B+}, A_d, \omega_d)\simeq ({k\over {\cal
H}_{B+}})^2$, which is consistent with (\ref{smallk}). $P_\zeta$
for the \textit{``Template"} (\ref{Template}) is plotted in Fig.
\ref{figmod5}. We see that (\ref{Template}) has effectively
captured the essential shape of $P_\zeta$ showed in Fig.
\ref{multi-power}.

\subsection{Data fitting}

We modified the CAMB and CosmoMC code package and perform a global
fitting with Planck2015 data. The parameter set of the
lensed-$\Lambda$CDM model is $\{\Omega_bh^2, \Omega_ch^2,
100\theta_\mathrm{MC}, \tau, \ln(10^{10}A_{inf}), n_{inf}\}$,
with $\Omega_bh^2$ the baryon density, $\Omega_ch^2$ the
cold dark matter density,  $\theta_\mathrm{MC}$ the angular size
of the sound horizon at decoupling, and  $\tau$ the reionization
optical depth. We also include the parameters set $\{{\cal
H}_{B+}, A_d, \omega_d\}$ (so-called the bounce
3-parameters) defined in (\ref{Template}), which captures the
physics of pre-inflationary bounce, as has been argued. We set the
pivot scale $k_{*}=0.05 Mpc^{-1}$, roughly in the middle of the
logarithmic range of scales probed by Planck.

With (\ref{Template}), we plot the CMB TT-spectrum 
$D_l^{TT}\equiv l(l+1)C_l^{TT}/2\pi$ and $\Delta D_l^{TT}$ in
Fig. \ref{figconstraint} with the best-fit parameters set
$\{\Omega_bh^2, \Omega_ch^2, 100\theta_\mathrm{MC}, \tau,
\ln(10^{10}A_{inf}), n_{inf}, {\cal H}_{B+}, A_d, \omega_d\}$.
Since WMAP and Planck, some models attempting to explain
the anomalies of CMB at large scale (but not solving the initial
singularity) have been proposed
\cite{Contaldi:2003zv}\cite{Dudas:2012vv}\cite{Das:2014ffa}\cite{Cai:2015xla}\cite{Wang:2016wio}\cite{Kontou:2017xhp}.
We see that the spectrum (\ref{Template}) of scalar perturbation
predicted by our model could fit better not only
the power deficit of the CMB TT-spectrum at low multipoles, but
also the dip at $l\sim 20$.  Actually, after we add the
bounce 3-parameters $\{{\cal H}_{B+}, A_d, \omega_d\}$ into the
parameter set of the $\Lambda$CDM model, the corresponding
$\Delta\chi^2$ value can be greatly improved. The details will be
presented in upcoming work.

\section{Conclusion}

In bounce inflation scenario, the inflation is singularity-free
(past-complete).  However, its pathology-free model has
been still lacking. Here, we showed such a model. The nonsingular
bounce is implemented by applying ${\tilde P}(\phi,X)$, see
(\ref{tildeP}), which is ghost-free, while $c_s^2<0$ is dispelled
by $L_{\delta g^{00} R^{(3)}}$ \cite{Cai:2017dyi}.

We perform a full simulation for the evolution of scalar
perturbation, and find that the spectrum $P_\zeta$ has a
suppression at large scale $k\ll {\cal H}_{B+}$ but is flat (with
a damped oscillation) at small scale $k\gg {\cal H}_{B+}$, which
confirms the earlier results showed in
\cite{Piao:2003zm}\cite{Liu:2013kea} and is consistent with the
power deficit of the CMB TT-spectrum at low multipoles $l\lesssim
30$; but unexpectedly, $P_\zeta$ also shows itself one marked
lower valley at $k\gtrsim {\cal H}_{B+}$, though the depth is
model-dependent. We show that this lower valley actually provides
a better fit to the dip at $l\sim 20$ hinted by Planck
\cite{Ade:2015lrj}. Based on the simulation and the analytical
estimation for the perturbation spectrum, we also offer a
``\textit{Template}" of $P_\zeta$ (effectively capturing the
physics of bounce) to fit data.

\begin{figure}[htbp]
\includegraphics[scale=2,width=0.6\textwidth]{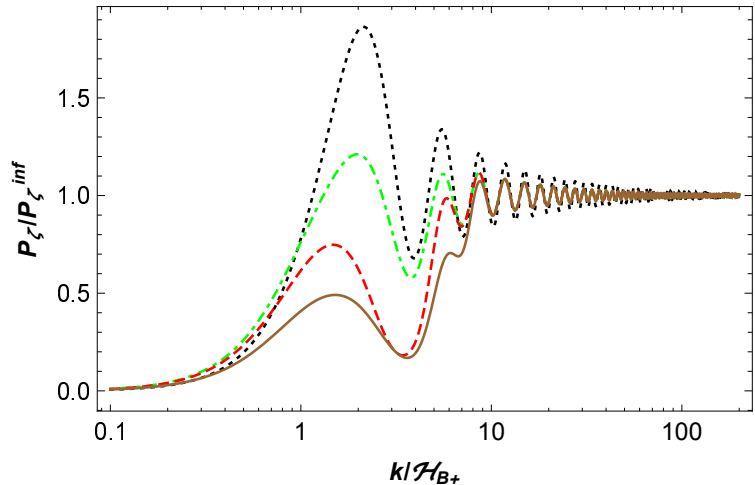}
\caption{The black dotted curve is the spectrum
$P_T/P_T^{inf}$ of the primordial GWs in bounce inflation
scenario, see \cite{Li:2016awk}, while the \{green dotdashed, red
dashed, brown solid\} curves are those of the primordial scalar
perturbation based on the results of \textit{``Template"}
(\ref{modfit3}) with $A_d=\{0.25,\,0.8,\,0.8\}$,
$d=\{\pi,\,\pi,\,\pi\}$ and $\omega_d=\{0.25,\,0.25,\,0.1\}$,
which are consistent with those in Fig. \ref{multi-power}. }
\label{figmod5}
\end{figure}

\begin{figure}[htbp]
    \subfigure[~~$l(l+1)C_l^{TT}/2\pi$]{\includegraphics[width=.44\textwidth]{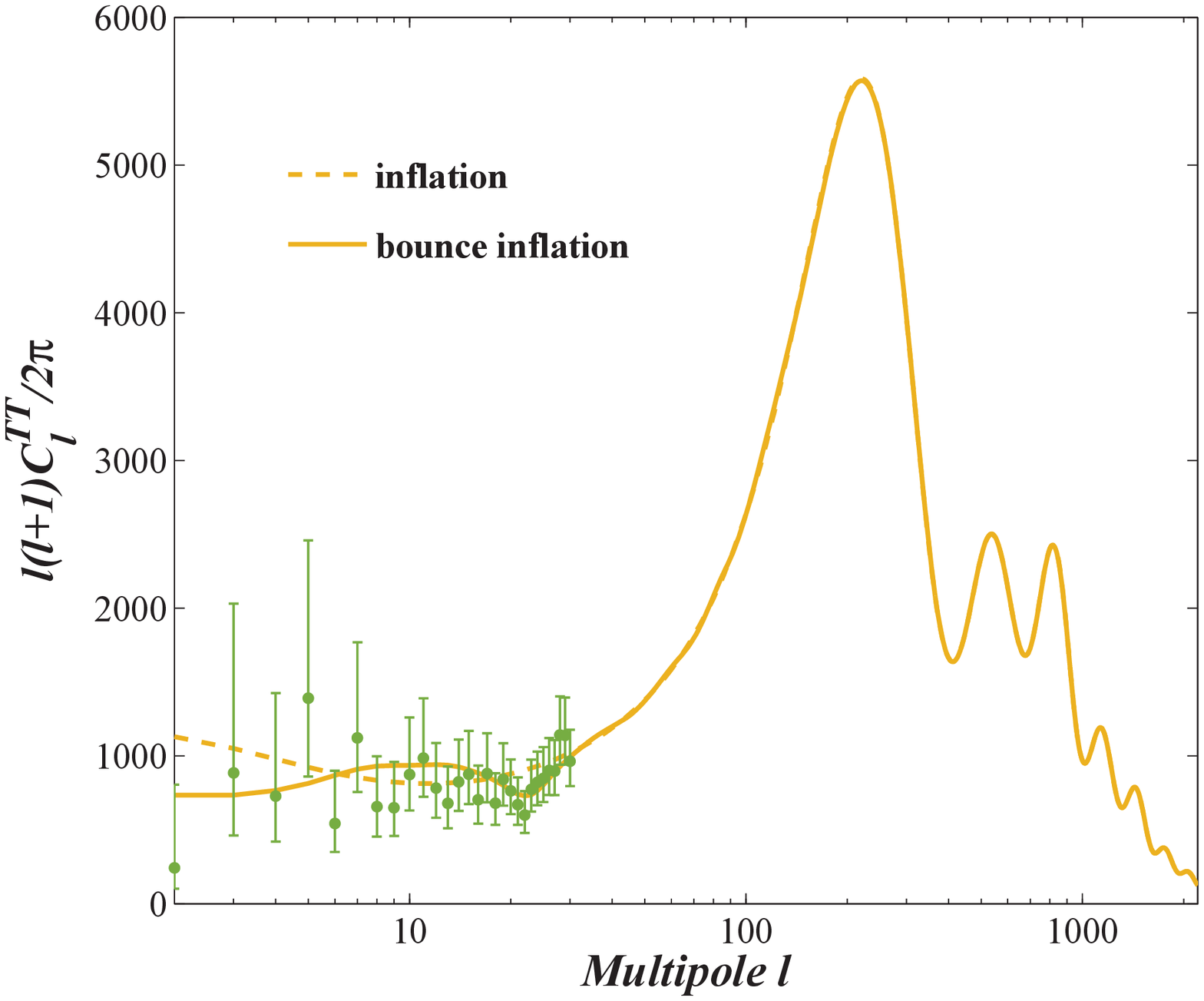} }
    \subfigure[~~$\Delta D_l^{TT}$]{\includegraphics[width=.47\textwidth]{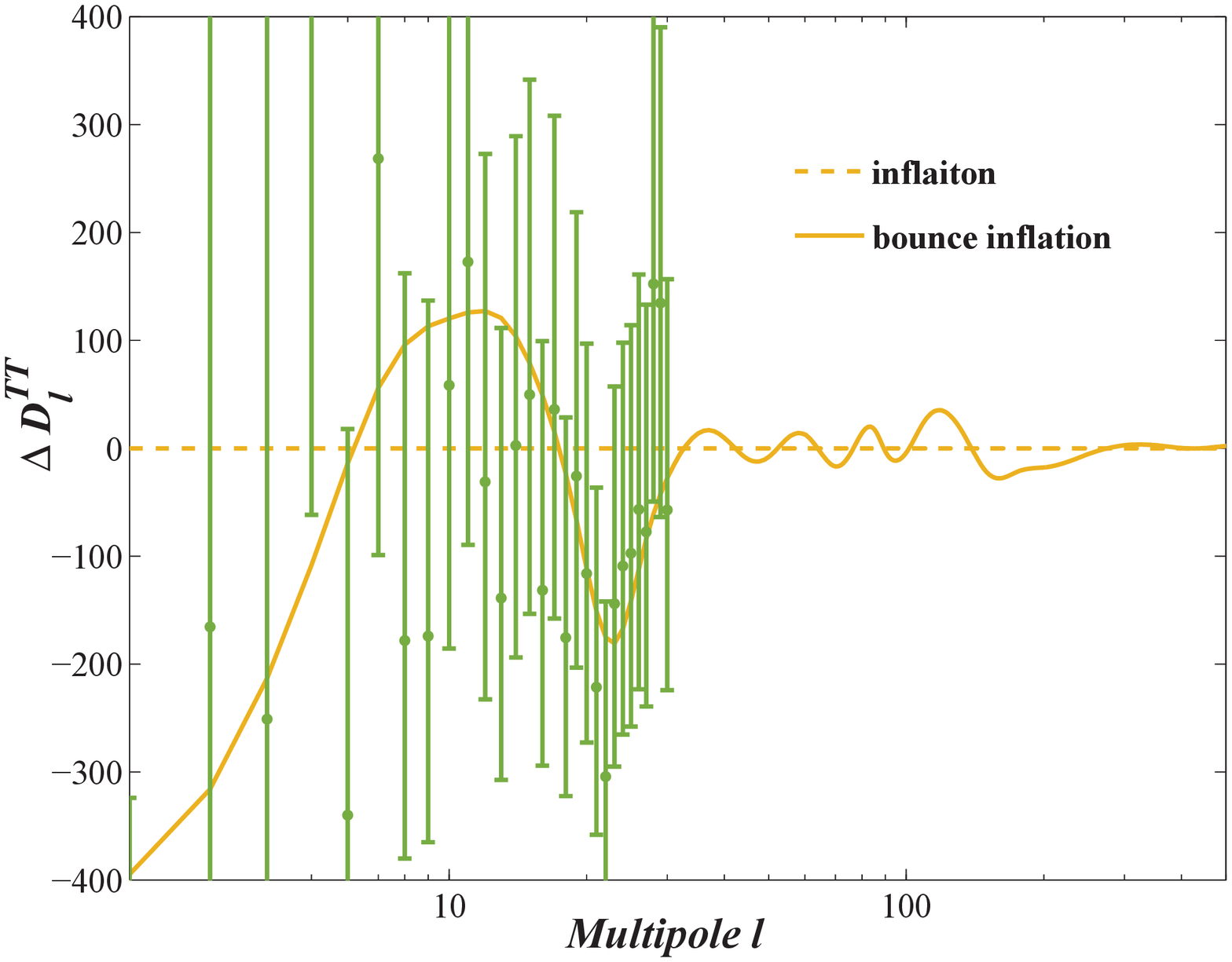} }
    \caption{The green points show the Planck2015 data with $1\sigma$
        errors. The best-fit values of parameters are
        $\ln(10^{10}A_{inf})$ = 3.091, $n_{inf}$ = 0.966, $\ln({\cal
            H}_{B+}) = -7.51$, $A_d = 0.87$, $\omega_d = 5.47$.} \label{figconstraint}
\end{figure}

The equation of motion  of GWs mode $\gamma_{ij}$ for
(\ref{action01}) is \be \ddot{\gamma}_k+\lf(3H+{{\dot
Q}_T\over Q_T}\rt)\dot{\gamma}_k+c_T^2{k^2\over
a^2}\gamma_k=0\,,\label{eom-gamma} \ee which is unaffected by the
operators $ R^{(3)}\delta g^{00}$ and $\delta K \delta g^{00}$,
where $Q_T=M_p^2$. We plot the primordial GWs spectrum $P_T$ in
Fig. \ref{figmod5} (the black dot curve) with
$P_T^{inf}={2H_{inf}^2\over \pi^2 M_p^2}$, see also
\cite{Li:2016awk}. It should be mentioned that if $Q_T\neq M_p^2$
around the nonsingular bounce (the gravity is modified
completely), $P_T$ will be different. It is also possible
that the corresponding gravity has a large parity violation
\cite{Wang:2014abh}, which might be imprinted in CMB.

Our work highlight the conjecture again that the physics hinted by
the large-scale anomalies of CMB is related with the
pre-inflationary bounce. The nonsingular cosmological bounce also
has been implemented in some models of modified gravity
\cite{Yoshida:2017swb,Misonoh:2016btv,Giovannini:2016jkf,Banerjee:2016hom,Biswas:2011ar,Vasilic:2017myy,Li:2016xjb,Odintsov:2015zza,Hendi:2016tiy,Chinaglia:2017wim,Giovannini:2017ndw,Farnsworth:2017wzr},
see also \cite{Nojiri:2017ncd}\cite{Battefeld:2014uga} for
reviews. Confronting the corresponding models with the CMB
data will be interesting.

\textbf{Acknowledgments}

YC would like to thank Youping Wan and Yi-Fu Cai for
discussions and hospitalities during his visit at University of
Science and Technology of China. YSP thanks Mingzhe Li for helpful
suggestions in USTC-ICTS seminar. We acknowledge the use of CAMB
and CosmoMC. This work is supported by NSFC, No. 11575188,
11690021, and also supported by the Strategic Priority Research
Program of CAS, No. XDA04075000, XDB23010100.

\appendix

\section{The EFT of nonsingular cosmologies}\label{appA}

In this Appendix, we briefly review the EFT of nonsingular
cosmologies, see \cite{Cai:2016thi} for the details.

With the ADM $3+1$ decomposition, we have
\begin{equation}
g_{\mu\nu}=\left(
  \begin{array}{cc}
  N_kN^k-N^2 &  N_j\\
  N_i &  h_{ij}\\
  \end{array}
\right) \,,\qquad
g^{\mu\nu}=\left(
  \begin{array}{cc}
  -N^{-2} &  {N^j\over N^2}\\
  {N^i\over N^2} &  h^{ij}-{N^iN^j\over N^2}\\
  \end{array}
\right) \,,\qquad
\end{equation}
and $\sqrt{-g}=N\sqrt{h}$, where $N_i=h_{ij}N^j$.  The induced
metric on 3-dimensional hypersurface is
$h_{\mu\nu}=g_{\mu\nu}+n_{\mu}n_{\nu}$, where $n_{\mu}=n_0
(dt/dx^{\mu})=(-N,0,0,0)$, $n^{\nu}=g^{\mu\nu}n_{\mu} =({1/
N},-{N^i/ N})$ is orthogonal to the spacelike hypersurface, and
$n_{\mu}n^{\mu}=-1$. Thus
\begin{equation}
h_{\mu\nu}=\left(
  \begin{array}{cc}
  N_kN^k &  N_j\\
  N_i &  h_{ij}\\
  \end{array}
\right) \,,\qquad
h^{\mu\nu}=\left(
  \begin{array}{cc}
  0 &  0\\
  0 &  h^{ij}\\
  \end{array}
\right) \,.\qquad
\end{equation}

The EFT is \cite{Cai:2016thi} \ba \label{eft_action} S&=&\int
d^4x\sqrt{-g}\Big[ {M_p^2\over2} f(t)R-\Lambda(t)-c(t)g^{00}
\nn\\
&\,&+{M_2^4(t)\over2}(\delta g^{00})^2-{m_3^3(t)\over2}\delta
K\delta g^{00} -m_4^2(t)\lf( \delta K^2-\delta K_{\mu\nu}\delta
K^{\mu\nu} \rt) +{\tilde{m}_4^2(t)\over 2}R^{(3)}\delta g^{00}
\nn\\
&\,&-\bar{m}_4^2(t)\delta K^2+{\bar{m}_5(t)\over 2}R^{(3)}\delta K
+{\bar{\lambda}(t)\over2}(R^{(3)})^2+...
\nn\\
&\,& -{\tilde{\lambda}(t)\over
    M_p^2}\nabla_iR^{(3)}\nabla^iR^{(3)} +... \Big] \,,
\ea where $\delta g^{00}=g^{00}+1$, $R^{(3)}$ is the 3-dimensional
Ricci scalar, $K_{\mu\nu}=h_{\mu}^{\sigma}\nabla_{\sigma}n_{\nu}$
is the extrinsic curvature, $\delta
K_{\mu\nu}=K_{\mu\nu}-h_{\mu\nu}H$.

Here, we focus on building a stable model of bounce
inflation. We only consider the coefficients set $(f, c, \Lambda,
M_2, m_3, {\tilde m}_4)$, and set other coefficients in
(\ref{eft_action}) equal to 0. We always could set $f=1$, which
suggests $c(t)=-M_p^2{\dot H}$ and $c(t)+\Lambda(t)=3M_p^2H^2$.

As pointed out in Ref. \cite{Cai:2017dxl}, the $R^{(3)}\delta K$
operator in EFT could play similar role as $R^{(3)}\delta g^{00}$,
which we will consider elsewhere. Mapping (\ref{action01}) into
the EFT (\ref{eft_action}), we have $M_2^4(t)= X^2{\tilde
P}_{XX}$, $m_3^3(t)= -g_1(\phi)$ and $\tilde{m}_4^2=f_1(\phi)$.
Only with $(M_2, m_3, {\tilde m}_4)\neq 0$, the quadratic action
of scalar perturbation $\zeta$ is (see, e.g., our
\cite{Cai:2016thi}) \ba \label{SS} S^{(2)}_\zeta=\int
d^4x\,a^3Q_s\lf( \dot{\zeta}^2-c_s^2{(\partial
    \zeta)^2\over a^2}
\rt)\,, \ea where \ba &\,&Q_s=\frac{2
M_2^4}{\gamma^2}+\frac{3m_3^6}{ 4M_p^2{\gamma^2}}-\frac{\dot{H}
M_p^2}{\gamma^2}\,,
\\ &\,&
c^2_sQ_s={\dot{c}_3\over a}-M_p^2 \,
\\
&\,&c_3=\frac{a M_p^2}{\gamma}\lf(1+ {2{\tilde m}_4^2\over
    M_p^2}\rt)\,,
\ea where $\gamma = H-{m_3^3/(2M_p^2)}$. Only if $Q_s>0$ and
$c_s^2>0$, the nonsingular cosmological model is healthy. In
models with the operator $(\delta g^{00})^2$, $Q_s>0$ always can
be obtained, since $(\delta g^{00})^2$ contributes ${\dot
\zeta}^2$. While $c_s^2>0$ requires ${\dot{c}_3}>aM_p^2$, which is
\be c_3|_{t_f}-c_3|_{t_i}>M_p^2\int_{t_i}^{t_f}\,adt\,.
\label{inequ}\ee The inequality (\ref{inequ}) suggests that $c_3$
must cross 0 (${\tilde m}_4^2=-M_p^2/2$ or $\gamma$ is divergent),
since the integral $\int adt$ is infinite. Thus if the
$R^{(3)}\delta g^{00}$ operator is absent, $c_s^2>0$ throughout is
impossible. We can set $c_s^2\simeq 1$ by \be \label{cs2eq1}
2{m}_4^2={\gamma\over a}\int a\lf(Q_s c_s^2+M_p^2\rt)dt-{M_p^2}\,.
\ee

\section{More on the simulation}\label{appB}

We plot the evolutions of $Q_s$, $g_1$, $|\zeta|$  with
respect to $t$, and also $P_\zeta(k)$ for the
background in Fig. \ref{fig01}, with different values of ${\cal
B}$ and $t_*$ in this Appendix.

\begin{figure}[htbp]
    \subfigure[~~$Q_s$]{\includegraphics[width=.47\textwidth]{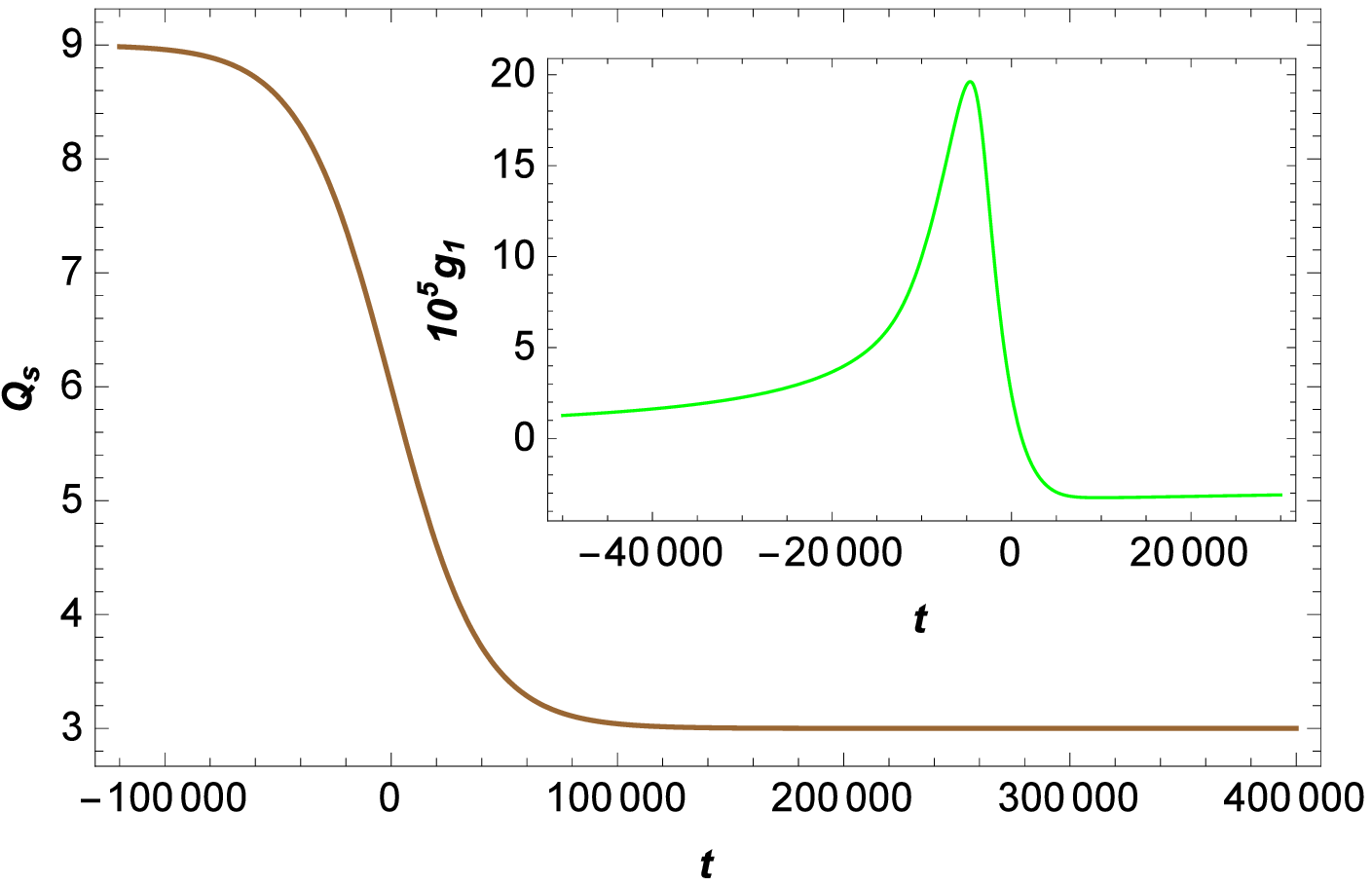} }
    \subfigure[~~$10^5\times(3H+\dot{Q}_s/Q_s)$]{\includegraphics[width=.47\textwidth]{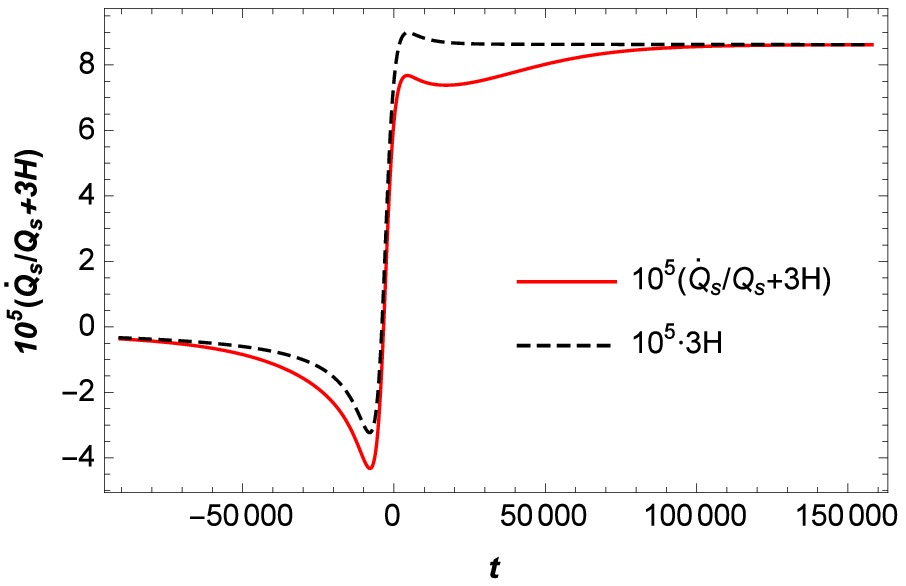} }
    \subfigure[~~$|\zeta|$ for $k=\{10^{-5}, 3\times10^{-5}, 10^{-3}, 10^{-2}\}$ from top to bottom ]{\includegraphics[width=.48\textwidth]{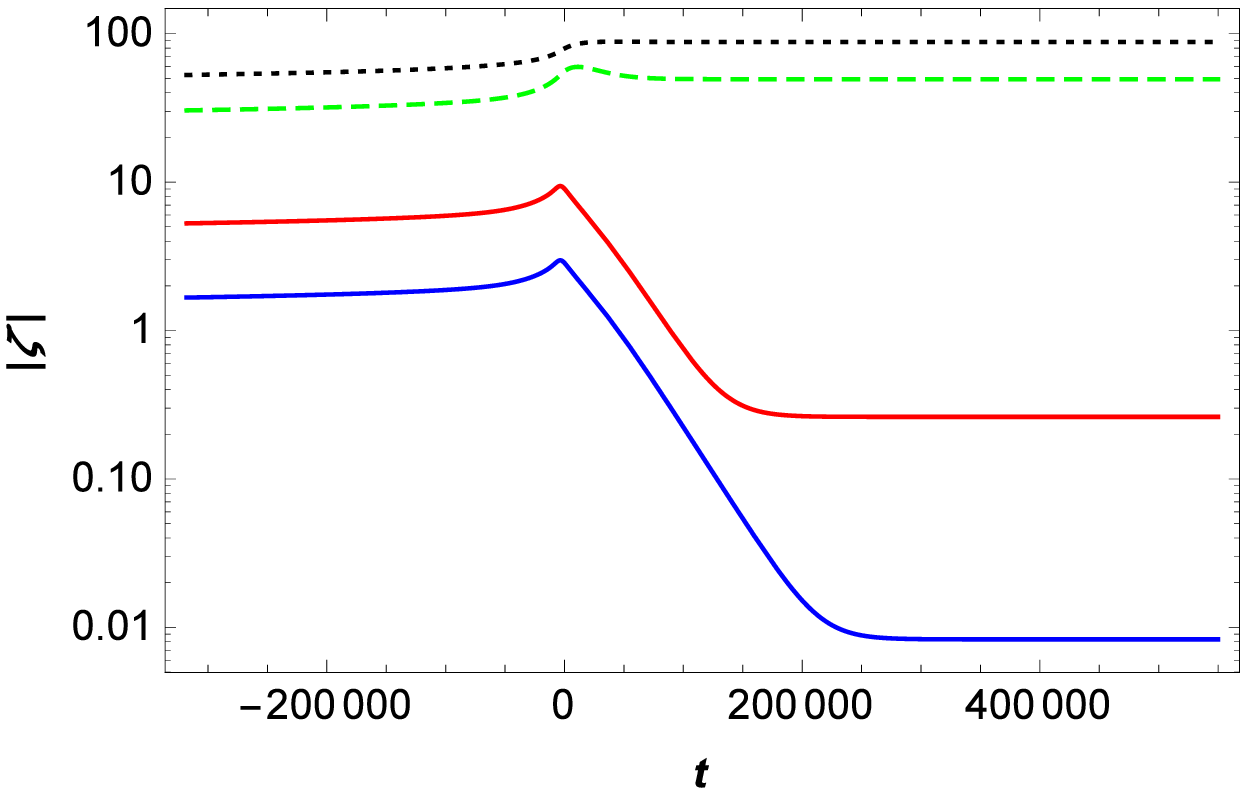} }
    \subfigure[~~$P_{\zeta}/P_{\zeta}^{inf}$]{\includegraphics[width=.47\textwidth]{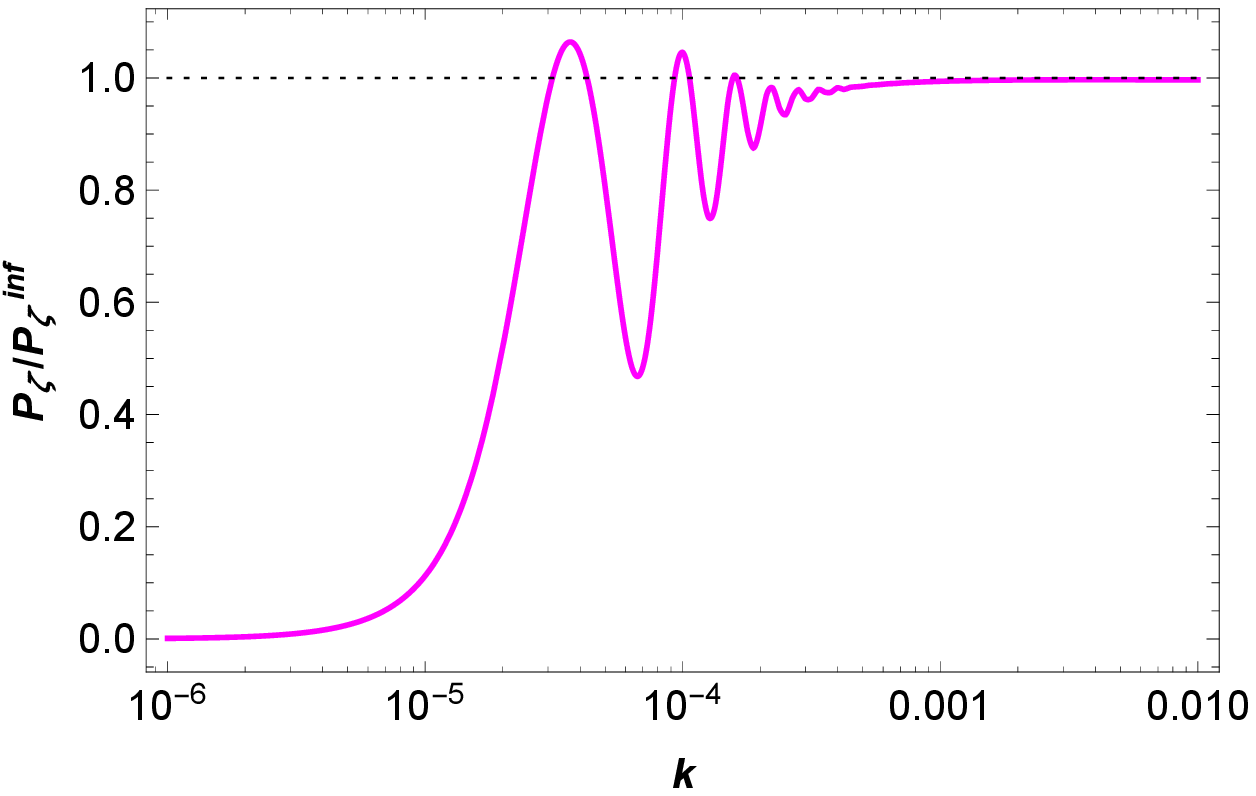} }
    \caption{We set ${\cal A}_Q=3$, ${\cal B}=2$, $t_*=4\times 10^4$ and the background is given by Fig. \ref{fig01}.} \label{fig02}
\end{figure}

\begin{figure}[htbp]
    \subfigure[~~$Q_s$]{\includegraphics[width=.47\textwidth]{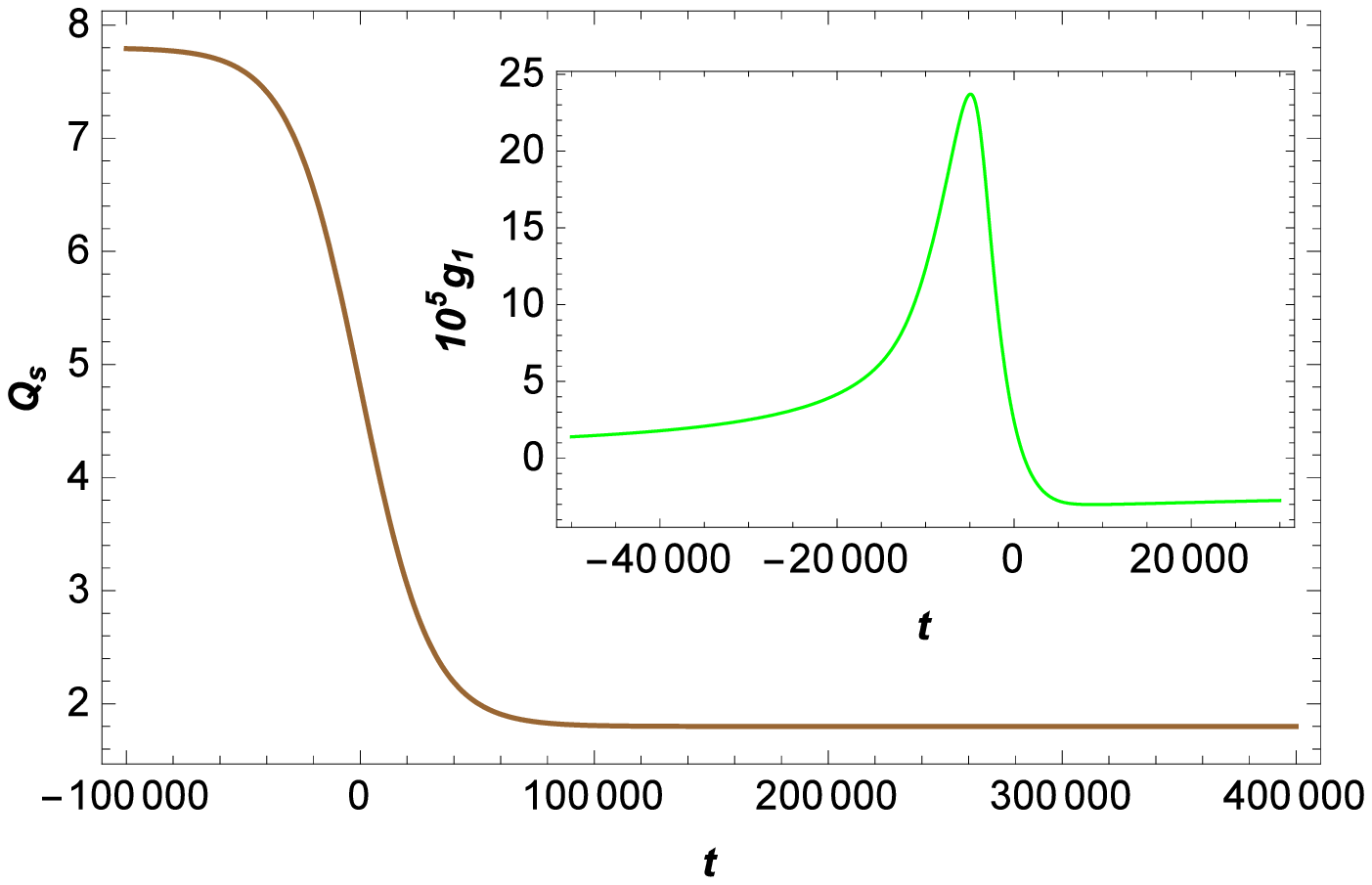} }
    \subfigure[~~$10^5\times(3H+\dot{Q}_s/Q_s)$]{\includegraphics[width=.47\textwidth]{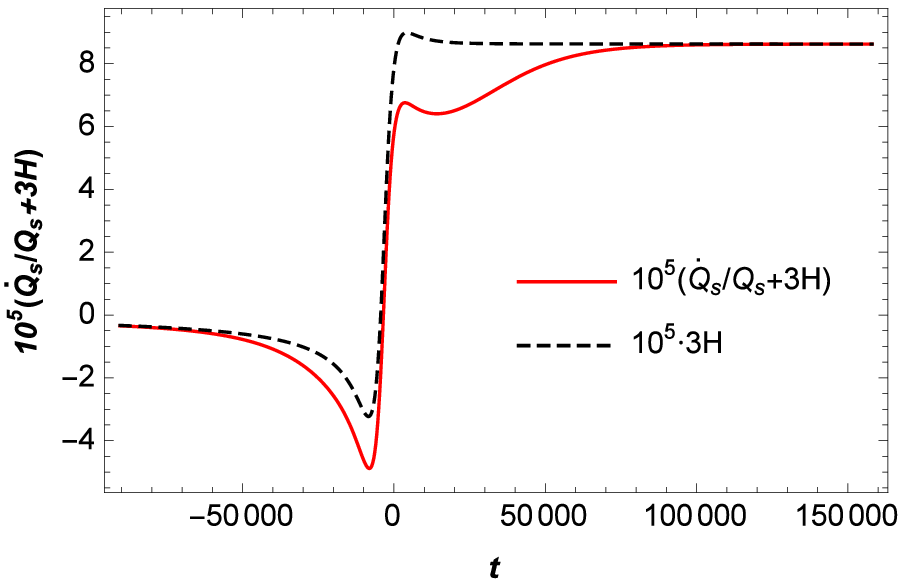} }
    \subfigure[~~$|\zeta|$ for $k=\{10^{-5}, 3\times10^{-5}, 10^{-3}, 10^{-2}\}$ from top to bottom ]{\includegraphics[width=.48\textwidth]{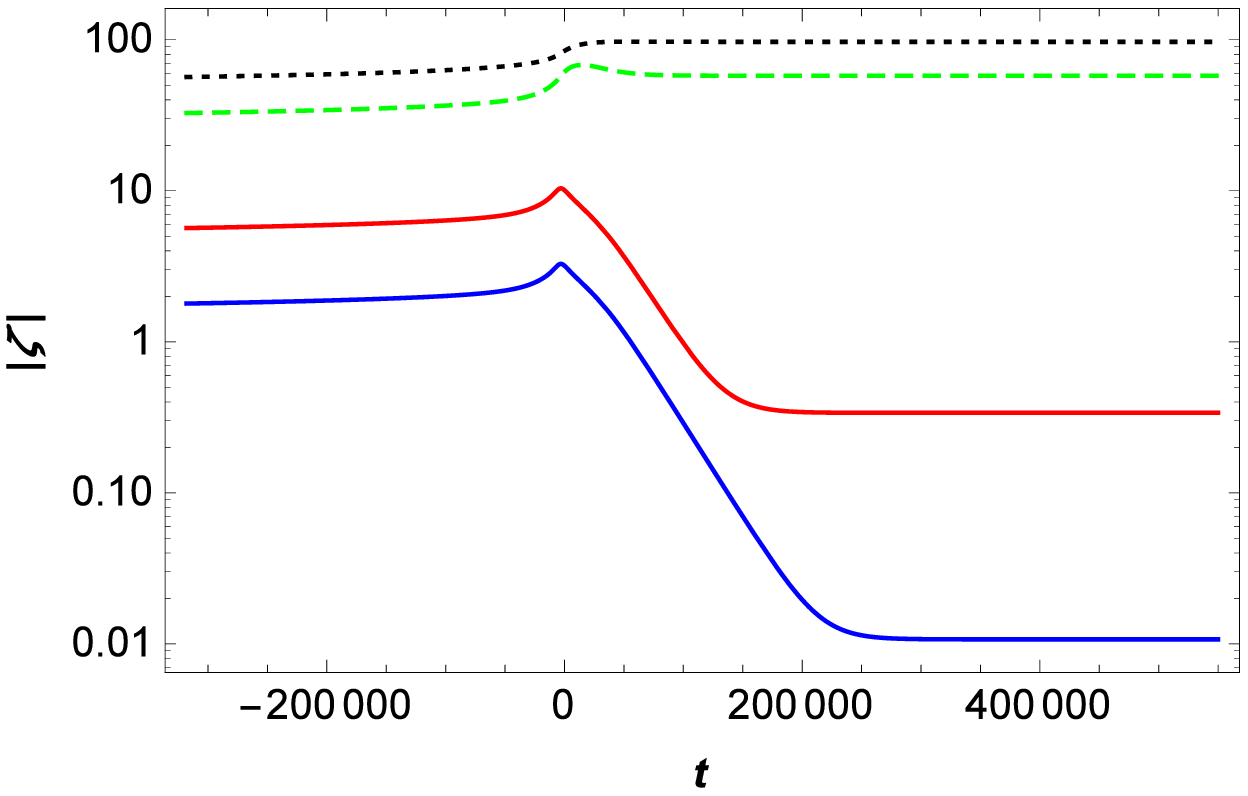} }
    \subfigure[~~$P_{\zeta}/P_{\zeta}^{inf}$]{\includegraphics[width=.47\textwidth]{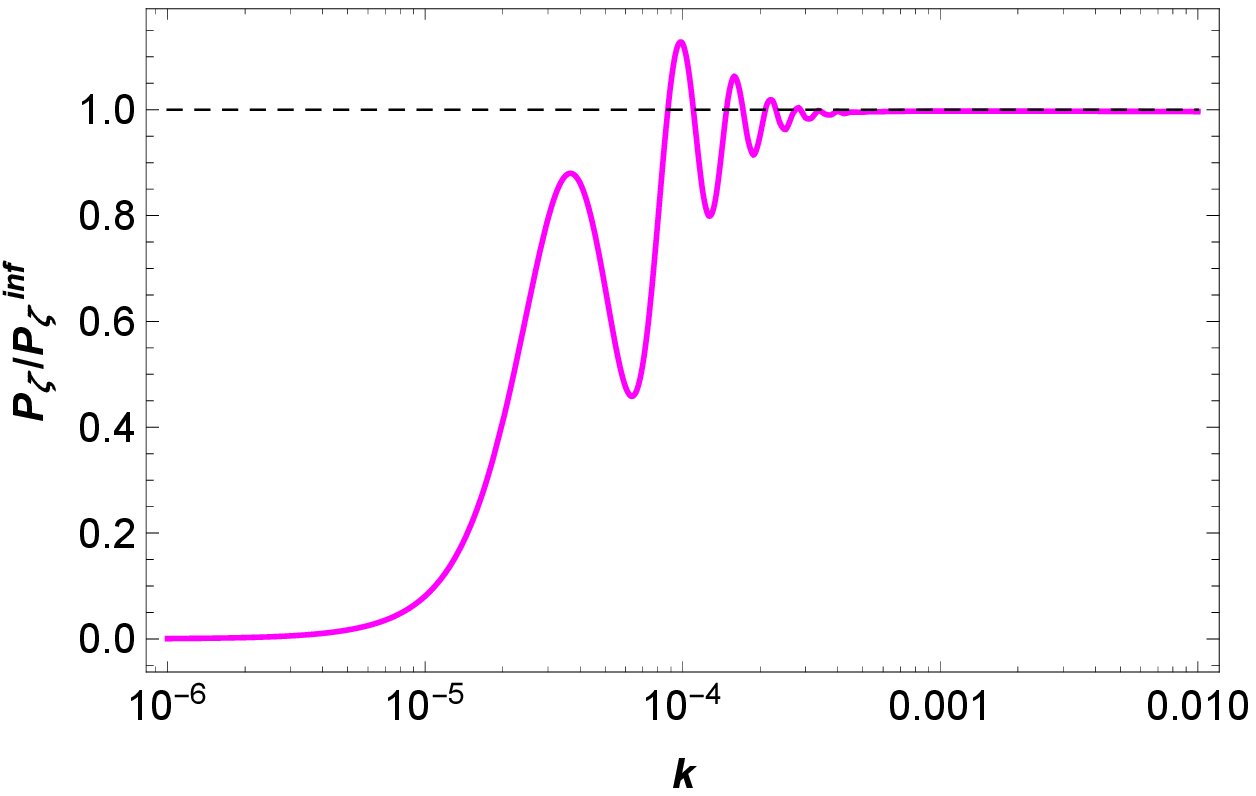} }
    \caption{We set ${\cal A}_Q=3$, ${\cal B}=1.6$, $t_*=3\times 10^4$
    and the background is given by Fig. \ref{fig01}.} \label{fig03}
\end{figure}

We see how
$|\zeta|$ evolves with $a$ in different phases.
Theoretically, $\zeta\sim 1/a$ for the perturbation modes
with $k\gg \sqrt{z_s''/z_s}$, while $\zeta\sim const.$ for the
perturbation modes with $k\ll \sqrt{z_s''/z_s}$, which is consistent with
our Figs. \ref{fig02}(c) and \ref{fig03}(c).

\section{The matrices elements of ${\cal{M}}^{(2,1)}$ and ${\cal M}^{(3,2)}$}\label{appC}

We define $l=\sqrt{{\alpha}{ a}_B^2-k^2}$, $x_1=1/|{\cal H}_{B-}|$,
$x_2={\cal H}_{B+}$,
$y_{1,2}=(\eta_{B{\mp}}-\eta_{B})$, and have
\ba {\cal{M}}^{(2,1)}_{11} &=& {\sqrt{\pi
x_1}\over4l}\lf[(l+{\alpha}{
a}_B^2y_1)H^{(1)}_{\nu_1}(kx_1)-kH^{(1)}_{\nu_1-1}(kx_1)\rt]e^{-ly_1},
\\
{\cal{M}}^{(2,1)}_{12} &=& {\sqrt{\pi x_1}\over4l}\lf[(l+{\alpha}{
a}_B^2y_1)H^{(2)}_{\nu_1}(kx_1)-kH^{(2)}_{\nu_1-1}(kx_1)\rt]e^{-ly_1},
\\
{\cal{M}}^{(2,1)}_{21}&=&{\sqrt{\pi x_1}\over4l}\lf[(l-{\alpha}{
a}_B^2y_1)H^{(1)}_{\nu_1}(kx_1)-kH^{(1)}_{\nu_1-1}(kx_1)\rt]e^{ly_1},
\\
{\cal{M}}^{(2,1)}_{22}&=&{\sqrt{\pi x_1}\over4l}\lf[(l-{\alpha}{
a}_B^2y_1)H^{(2)}_{\nu_1}(kx_1)-kH^{(2)}_{\nu_1-1}(kx_1)\rt]e^{ly_1},
\ea \ba {\cal{M}}^{(3,2)}_{11}&=&{i\sqrt{\pi
x_2}\over2}\lf[(l-{\alpha}{
a}_B^2y_2)H^{(2)}_{\nu_2}(kx_2)+kH^{(2)}_{\nu_2-1}(kx_2)\rt]e^{ly_2},
\\
{\cal{M}}^{(3,2)}_{12}&=&{i\sqrt{\pi x_2}\over2}\lf[-(l+{\alpha}{
a}_B^2y_2)H^{(2)}_{\nu_2}(kx_2)+kH^{(2)}_{\nu_2-1}(kx_2)\rt]e^{-ly_2},
\\
-{\cal{M}}^{(3,2)}_{21}&=&{i\sqrt{\pi x_2}\over2}\lf[(l-{\alpha}{
a}_B^2y_2)H^{(1)}_{\nu_2}(kx_2)+kH^{(1)}_{\nu_2-1}(kx_2)\rt]e^{ly_2},
\\
-{\cal{M}}^{(3,2)}_{22}&=&{i\sqrt{\pi x_2}\over2}\lf[-(l+{\alpha}{
a}_B^2y_2)H^{(1)}_{\nu_2}(kx_2)+kH^{(1)}_{\nu_2-1}(kx_2)\rt]e^{-ly_2}.
\ea

 \end{document}